\documentclass[12pt,onecolumn,twoside]{IEEEtran}

\usepackage{calc}

\usepackage{multirow}
\usepackage{cite}
\usepackage{graphicx,subfigure}
\usepackage{psfrag}
\usepackage{amsmath,amssymb,amsthm}
\usepackage{color}
\usepackage{tikz} 
\usepackage{comment}

\usepackage{cases}

\usepackage{algorithm}
\usepackage{algpseudocode}
\usepackage{mathtools}

\usepackage{pgfplots}
\usepackage{scalefnt}

\DeclarePairedDelimiter\ceil{\lceil}{\rceil}
\DeclarePairedDelimiter\floor{\lfloor}{\rfloor}
\makeatletter
\def\BState{\State\hskip-\ALG@thistlm}
\makeatother

\algnewcommand\algorithmicswitch{\textbf{switch}}
\algnewcommand\algorithmiccase{\textbf{case}}
\algnewcommand\algorithmicassert{\texttt{assert}}
\algnewcommand\Assert[1]{\State \algorithmicassert(#1)}%
\algdef{SE}[SWITCH]{Switch}{EndSwitch}[1]{\algorithmicswitch\ #1\ \algorithmicdo}{\algorithmicend\ \algorithmicswitch}%
\algdef{SE}[CASE]{Case}{EndCase}[1]{\algorithmiccase\ #1}{\algorithmicend\ \algorithmiccase}%
\algtext*{EndSwitch}%
\algtext*{EndCase}%

\algnewcommand\algorithmicforeach{\textbf{for each}}
\algdef{S}[FOR]{ForEach}[1]{\algorithmicforeach\ #1\ \algorithmicdo}
\renewcommand\algorithmicdo{}

\DeclareMathOperator*{\defeq}{\triangleq}

\newtheorem{theorem}{Theorem}

\newcommand{\bit}{\begin{itemize}}
\newcommand{\eit}{\end{itemize}}

\newcommand{\bc}{\begin{center}}
\newcommand{\ec}{\end{center}}

\newcommand{\ba}{\begin{array}}
\newcommand{\ea}{\end{array}}

\newcommand{\beq}{\begin{equation}}
\newcommand{\eeq}{\end{equation}}

\newcommand{\beqn}{\begin{equation*}}
\newcommand{\eeqn}{\end{equation*}}

\newcommand{\bean}{\begin{eqnarray*}}
\newcommand{\eean}{\end{eqnarray*}}
\newcommand{\bea}{\begin{eqnarray}}
\newcommand{\eea}{\end{eqnarray}}

\def\E{\mathbb{E}}

\def\F{\mathbb{F}}

\def\av{\boldsymbol{a}}

\def\ev{\boldsymbol{e}}

\def\hv{\boldsymbol{h}}

\def\sv{\boldsymbol{s}}

\def\vv{\boldsymbol{v}}

\def\xv{\boldsymbol{x}}
\def\yv{\boldsymbol{y}}
\def\zv{\boldsymbol{z}}

\def\Hm{\boldsymbol{H}}

\def\Vm{\boldsymbol{V}}

\newcommand{\Dc}{{\mathcal D}}

\newcommand{\Gc}{{\mathcal G}}

\newcommand{\Jc}{{\mathcal J}}

\newcommand{\Lc}{{\mathcal L}}
\newcommand{\Mc}{{\mathcal M}}

\newcommand{\Pc}{{\mathcal P}}

\newcommand{\Rc}{{\mathcal R}}
\newcommand{\Sc}{{\mathcal S}}

\newcommand{\Wc}{{\mathcal W}}

\newtheorem{remark}{Remark}

\newcommand{\non}{\nonumber}

\usepackage{tabu}
\usepackage{mathtools}
\usepackage{xifthen}
\usepackage{booktabs}

\newcommand{\bunderline}[1]{\underline{#1\mkern-4mu}\mkern4mu }

\newcommand{\hvu}{\bunderline{\boldsymbol{h}}}
\newcommand{\xvu}{\bunderline{\boldsymbol{x}}}
\newcommand{\uvu}{\bunderline{\boldsymbol{u}}}
\newcommand{\vvu}{\bunderline{\boldsymbol{v}}}
\newcommand{\yvu}{\bunderline{\boldsymbol{y}}}
\newcommand{\zvu}{\bunderline{\boldsymbol{z}}}
\newcommand{\svu}{\bunderline{\boldsymbol{s}}}
\newcommand{\qvu}{\bunderline{\boldsymbol{q}}}

\newcommand{\fvu}{\bunderline{\boldsymbol{f}}}

\newcommand{\xvut}[1][]{\ifthenelse{\isempty{#1}}{\xvu_{t}}{\xvu_{#1}}}
\newcommand{\hvut}[1][]{\ifthenelse{\isempty{#1}}{\hvu_{t}}{\hvu_{#1}}}
\newcommand{\uvut}[1][]{\ifthenelse{\isempty{#1}}{\uvu_{t}}{\uvu_{#1}}}
\newcommand{\vvut}[1][]{\ifthenelse{\isempty{#1}}{\vvu_{t}}{\vvu_{#1}}}
\newcommand{\svut}[1][]{\ifthenelse{\isempty{#1}}{\svu_{t}}{\svu_{#1}}}
\newcommand{\qvut}[1][]{\ifthenelse{\isempty{#1}}{\qvu_{t}}{\qvu_{#1}}}
\newcommand{\zvut}[1][]{\ifthenelse{\isempty{#1}}{\zvu_{t}}{\zvu_{#1}}}
\newcommand{\yvut}[1][]{\ifthenelse{\isempty{#1}}{\yvu_{t}}{\yvu_{#1}}}
\newcommand{\fvut}[1][]{\ifthenelse{\isempty{#1}}{\fvu_{t}}{\fvu_{#1}}}

\newcommand{\hvt}[1][]{\ifthenelse{\isempty{#1}}{\hv_{t,m}}{\hv_{t,#1}}}
\newcommand{\hvot}[1][]{\ifthenelse{\isempty{#1}}{\hv_{1,m}}{\hv_{1,#1}}}
\newcommand{\zt}[1][]{\ifthenelse{\isempty{#1}}{\zv_{t}}{\zv_{#1}}}
\newcommand{\yt}[1][]{\ifthenelse{\isempty{#1}}{\yv_{t}}{\yv_{#1}}}
\newcommand{\st}[1][]{\ifthenelse{\isempty{#1}}{\sv_{t}}{\sv_{#1}}}

\begin{document}
\sloppy

\title{Wireless MapReduce Distributed Computing}

\author{Fan Li, Jinyuan Chen and Zhiying Wang

\thanks{Fan Li and Jinyuan Chen  are with Louisiana Tech University, Department of Electrical Engineering, Ruston, LA 71272, US (emails: fli005@latech.edu, jinyuan@latech.edu).   Zhiying Wang is with University of California,  Irvine, Center for Pervasive Communications and Computing (CPCC),  Irvine, CA 92697, US  (email:zhiying@uci.edu). This work was presented in part at the 2018 IEEE International Symposium on Information Theory.}}

\maketitle
\thispagestyle{empty}

\begin{abstract}
Motivated by mobile edge computing and wireless data centers, we study a wireless distributed computing framework where the distributed nodes exchange information over a wireless interference network. Our framework follows the structure of MapReduce. This framework consists of Map, Shuffle, and Reduce phases, where Map and Reduce are computation phases and Shuffle is a data transmission phase. In our setting, we assume that the transmission is operated over a wireless interference network. We demonstrate that, by duplicating the computation work at a cluster of distributed nodes in the Map phase, one can reduce the amount of transmission load required for the Shuffle phase. In this work, we characterize the fundamental tradeoff between computation load and communication load, under the assumption of one-shot linear schemes. The proposed scheme is based on side information cancellation and zero-forcing, and 
we prove that it is optimal in terms of computation-communication tradeoff. The proposed scheme outperforms the naive TDMA scheme with single node transmission at a time, as well as the coded TDMA scheme  that  allows coding across data, in terms of the  computation-communication tradeoff.

\end{abstract}

\section{Introduction}

In recent years, communication continuously moves from wireline to wireless links. For example, traffic from wireless and mobile devices will account for two-thirds of total IP traffic by 2020 (cf.~\cite{index2017zettabyte}). 
For another example, wireless data centers (e.g. \cite{VRBP:14, BDZLWG:15}) have become  attracting solutions due to the low cost for cabling. 
Moreover, distributed computing is popular for its capability of processing  a large amount of data in distributed nodes. The applications include mobile edge computing where the computing nodes are distributed mobile devices, as well as fog computing for Internet of things (IoT) with distributed computing nodes.
In this work, we study MapReduce distributed computing over a  wireless interference network.

In MapReduce distributed computing (cf.~\cite{DGmapreduce:2004,  LMA:15, LMYA:17}), data is first split and processed (called \emph{Map}) at the distributed nodes, and then the results are \emph{shuffled} (called \emph{Shuffle}), and  processed again (called \emph{Reduce}). As the amount of data and the number of nodes grow, the Shuffle phase could lead to a significant delay for the overall performance. In this work, we study a MapReduce-based wireless distributed computing framework, where the Shuffle phase is operated over a  wireless interference network, and explore the advantages of wireless communication  to reduce the system latency.

\begin{figure}
\centering
\includegraphics[width=13cm]{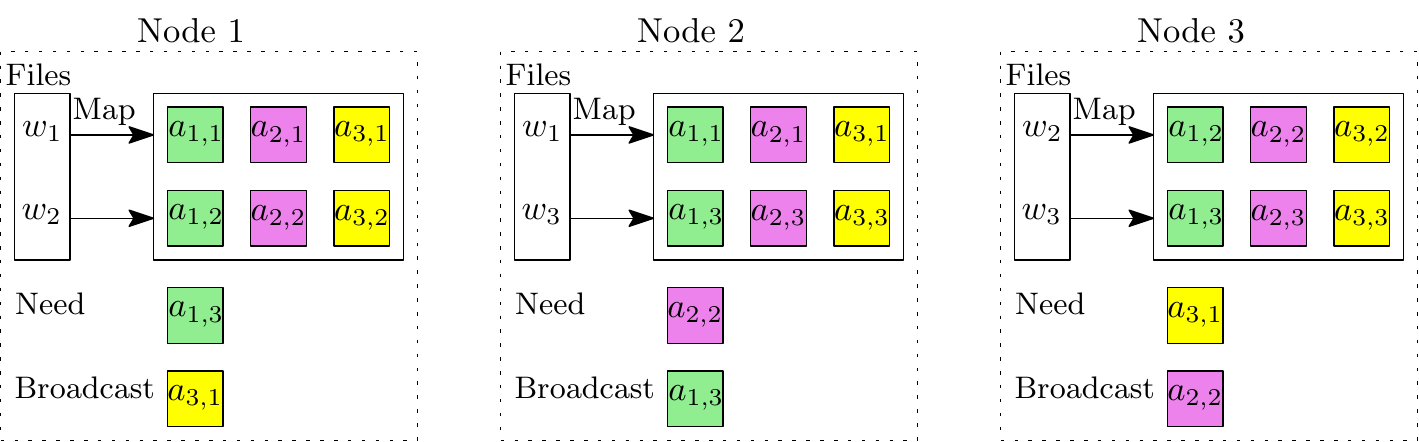}
\caption{An example of wireless distributed computing with $K=Q=N=3$ and $r=2$.}
\label{fig:KQN3}
\end{figure}

We parameterize the MapReduce problem  by $N,K,r,Q$, where $N$ is the number of data files,
$K$ is the number of nodes, each file is duplicated at  $r$  nodes on average (called \emph{computation load}),  
and $Q$ is the number of Reduce functions.
See Fig.~\ref{fig:KQN3} for an example. 
In this example, three  distributed nodes ($K=3$) seek to compute three  Reduce functions ($Q=3$) for  three  data files ($N=3$),  with each file stored at two nodes ($r=2$).
Every Map function takes one file as input, and outputs $3$ intermediate values, one for each Reduce function.
The intermediate value is denoted as $a_{q,n}$ for File~$n$ and Reduce function $q$.
The Reduce function $q$ takes $(a_{q,1}, a_{q,2},a_{q,3})$ as inputs and produces the $q$-th final value.
In the Map phase, every node  computes 6 intermediate values for 2 files. 
For example, Node 1 computes 6 intermediate values, i.e., $\{a_{q,n}: q=1,2,3, n=1,2\}$, for Files 1 and 2. 
In the Shuffle phase, some intermediate values are communicated in order to complete the computation in the Reduce phase. In the Reduce phase, assume that Node $k$  computes the $k$-th Reduce function, for $k=1,2,3$.
In order to compute the first Reduce function, Node 1 needs input $(a_{1,1}, a_{1,2}, a_{1,3})$. While $a_{1,1}$ and $a_{1,2}$ are already cached locally, $a_{1,3}$ needs to be transmitted from a different node in the Shuffle phase. Similarly, Node 2 requires $a_{2,2}$ and Node 3 requires $a_{3,1}$ in the Shuffle phase.

In our setting,  communication in the Shuffle phase takes place over a wireless interference channel. 
Assume that the channel state information is available to all nodes, and the communication is full-duplex. 
One possible application scenario is in data centers, where the environment (and hence the channel) is fixed for a long enough period, hence one may assume that channel state information is available at all users.
Let the (non-interfered) transmission time of 1 intermediate value be 1 time unit, namely, a coded packet corresponding to $a_{q,n}$ is transmitted using 1 time unit, such that $a_{q,n}$ can be successfully decoded.
 In order to handle interference, we have the following possible solutions.
\begin{itemize}
\item
If we use a naive \emph{uncoded} time-division multiple access (TDMA) broadcast scheme, allowing only 1 node to transmit 1 intermediate value at any time unit, we need 3 time units to transmit in total.  
\item
We could also use a \emph{coded} TDMA broadcast scheme (cf.~\cite{LMYA:17}), allowing only 1 node to transmit 1 \emph{coded} intermediate value at any time. For example, Node 3 can transmit a linear combination of the coded packets of $a_{1,3}$ and  $a_{2,2}$. Through the cached intermediate values, Nodes 1 and 2 can  respectively decode their desired information. Then Node 1 can transmit $a_{3,1}$ for Node 3. We need 2 time units in total.  
\item
Alternatively, we can let 3 nodes transmit \emph{at the same time}. Each node receives the superposition of the 3 transmitted symbols. However, the two undesired symbols can be canceled using  cached intermediate values (side information). Thus the desired symbol is decoded. We need only 1 time unit.
\end{itemize}

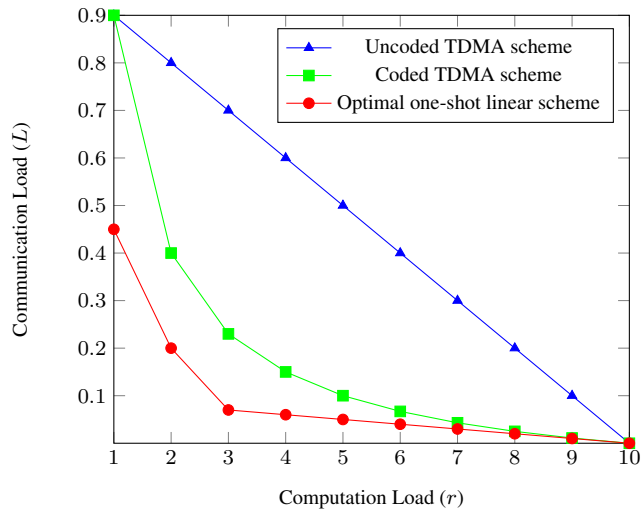
\begin{figure}
\begin{center}
{\scalefont{0.65}
\begin{tikzpicture}[scale = 1.0]
\begin{axis}[ 
        xmin=1, xmax=10,
	   ymin=0, ymax=0.9,
    	   xtick={1,2,3,4,5,6,7,8,9,10},
	   ytick={0.1,0.2,0.3,0.4,0.5,0.6,0.7,0.8,0.9},
    	legend pos=north east,
	xlabel=Computation Load ($r$),
	ylabel=Communication Load ($L$)]
\addplot[color=blue,mark=triangle*]
 coordinates {
	(1, 0.9)
	(2, 0.8)
	(3, 0.7)
	(4, 0.6)
	(5, 0.5)
	(6, 0.4)
	(7, 0.3)
	(8, 0.2)
	(9, 0.1)
     (10, 0 )
};

\addplot[color=green,mark=square*] 
	coordinates {
	(1, 0.9)
	(2, 0.4)
	(3, 0.23)
	(4, 0.15)
	(5, 0.10)
	(6, 0.067)
	(7, 0.043)
	(8, 0.025)
	(9, 0.011)
     (10, 0 )
};
\addplot[color=red,mark=*] 
	coordinates {
	(1, 0.45)
	(2, 0.2)
	(3, 0.07)
	(4, 0.06)
	(5, 0.05)
	(6, 0.04)
	(7, 0.03)
	(8, 0.02)
	(9, 0.01)
     (10, 0 )
};

\legend{Uncoded TDMA scheme, Coded TDMA scheme, Optimal one-shot linear scheme}
\end{axis}
\end{tikzpicture}
\vspace{-9pt}
\caption{Comparison on the communication load vs. computation load performance for uncoded TDMA scheme, coded TDMA, and the optimal one-shot linear scheme, given $K=10, N=2520$, and $Q=360$.} \label{fig:results}
} 
\end{center}
\end{figure}

In this paper we study the shuffle communication time units normalized by $NQ$, termed as \emph{communication load}, which is a function of $K$ and the computation load $r$. For practical purposes, we assume that the \emph{one-shot linear} scheme is used, where each intermediate value is encoded into a coded packet, and the transmitted symbol is a linear combination of the coded packets in the cache, ensuring that the coded packet can be decoded at the intended receiver with a linear operation.
We show that the optimal communication load is given as 
\begin{align}
\frac{1-\frac{r}{K}}{\min\{K,2r\}}, \quad r \in \{1,2,\dots,K\}. \label{eq1}
\end{align}
The significant improvement of our scheme compared to  uncoded and coded TDMA schemes is depicted in Fig.~\ref{fig:results}. As shown in Fig.~\ref{fig:results}, considering the  case of $r=1$, namely, when there is no extra computation in the Map phase,  the communication load of the proposed one-shot linear scheme is $50\%$ lower than that of both uncoded TDMA and coded TDMA schemes. For the case of $r=5$, the communication  load of the proposed one-shot linear scheme is $90\%$ lower than that of uncoded TDMA scheme and $50\%$ lower than that of coded TDMA scheme. 

The two key factors to obtain \eqref{eq1} are side information cancellation and zero-forcing. The role of side information has been demonstrated in the example of Fig.~\ref{fig:KQN3}. If an intermediate value is stored in multiple nodes, then by simultaneously transmitting this intermediate value from these nodes, the corresponding signal may be zero-forced at some undesired receivers. It is  similar to the interference cancellation in a MISO interference channel. In fact, we convert our problem to a MISO interference channel problem to obtain the converse. 

In this paper, the technical challenges lie in both the converse and achievability. 
For the converse, our main task is to bound the maximum number of coded packets that can be transmitted simultaneously at the $\ell$-th time unit, denoted by $|\Dc_{\ell}|$.  
When  each file is replicated $r$ times, referred to as  \emph{symmetric file replications},  we prove that $|\Dc_{\ell}|$  is upper bounded by a value that depends on the number of times each file is replicated, i.e., $r$.
However, when different files are replicated with different numbers of times, referred to as \emph{asymmetric file replications}, the problem becomes  more challenging, because we have $N$ parameters, each  corresponding to the replication number of one file.
For this case, even though each $|\mathcal{D}_\ell|$ depends on the replication numbers of the particular files involved in time unit $\ell$, we prove that the total number of required transmission time units is upper bounded by a value that depends on the average number of times the files are replicated (i.e., $r$).  In fact, this proof combined with our achievability shows that asymmetric file replications  cannot have a better communication load than symmetric ones.

For the achievability, we provide an explicit  one-shot linear  scheme,  in which files are placed symmetrically,  and the number of transmitted coded packets at each time unit attains the maximum of $|\mathcal{D}_{\ell}|$ from the converse. 
Note that  the difficulty of the achievability  lies in the case with $r<K/2$, where interference might not be eliminated completely if all nodes participate in transmission simultaneously. 
For this case, the proposed scheme guarantees that  a subset of nodes can receive packets
without interference at each  time unit, by using side information cancellation and \emph{partial} zero-forcing.

{\bf \emph{Related work:}}
In \cite{LMA:15, LMYA:17}, coded MapReduce was introduced to utilize cache and broadcast to reduce communication delay. 
A lot of work has appeared after that regarding communication in distributed computation, e.g. \cite{YLMA:17, EKF:17, LYMA:16, ATallerton:16, SFZ:17, AT:16, SLCarxiv:18, YYW:18, PLE:18, WCJ:18} and the references therein.  
Note that in another research direction of distributed computing, a number of works focused on mitigating the effect of stragglers and minimizing system latency  by using coding (cf.~\cite{DCG:16, DCG:17, LMAAllerton:16, Lee2015speeding, AT:17, LSR:17, KSD:17, RPPA:17, RP:17, TLDK:17, PLSSM:18}).
On the other hand, communication under wireless networks was studied for 
distributed computation (cf.~\cite{LYMAwireless:17} and \cite{LMAISIT:17}) and
content distribution (cf.~\cite{MNICC:15,HND:16, STS:17, JCM:16, NMA:17, xu2017fundamental}). 
Note that,  the setting of this paper is very different from the settings in both \cite{LYMAwireless:17} and \cite{LMAISIT:17}.  In the setting of \cite{LYMAwireless:17},  the distributed nodes must be connected through  a wireless access point (or a relay), while in the setting of this paper the distributed nodes  can communicate with each other and the communication channel is a wireless \emph{interference} channel.  In the setting of \cite{ LMAISIT:17}, mobile users ask the distributed computing nodes (helpers) to help compute the output functions, in the presence of a wireless communication network between the helpers and mobile users, while in the setting of this paper each node is a computing node. 
The coding approach considered here can also be applied to the other frameworks of distributed systems, for example,  federated learning, which is popular for distributed learning over distributed nodes \cite{KMYRSB:17, MMRHA:17, ASYKM:18}.

The remainder of this work is organized as follows. 
Section~\ref{sec:system} describes the system model. 
Section~\ref{sec:mainresult}  provides  the main results of this work. 
The  converse proof is described in Sections~\ref{sec:converse},  while the achievability  proof is described in  Sections~\ref{sec:ach}.  
Section~\ref{sec:example}  provides  the scheme examples. 
The work is concluded in Section~\ref{sec:conclusion}.

Throughout this work, $[c_1:  c_2 ]$ denotes the set of integers from $c_1$ to $c_2$, for some nonnegative integers $c_1 \le c_2$.  $|\bullet|$ denotes the magnitude of a scalar or the cardinality of a set.
$o(\bullet)$ comes from the standard Landau notation, where 
$f(x)=o(g(x))$ implies that $\lim_{x \to \infty} f(x)/g(x) =0$. 
 $\mathbb{C}$ and  $\mathbb{R}$ denote the sets of  complex numbers and real numbers, respectively.    
 $\F^{q}_{2}$ denotes the set of $q$-tuples over the binary field.    $\mathbb{N}^+$ denotes the set of positive natural numbers.
Logarithms are in base~$2$.   $\ceil*{c}$ denotes the least integer that is no less than $c$, and $\floor*{c}$ denotes the greatest integer that is no larger than $c$. 
$s \sim \mathcal{CN}(0, \sigma^2)$  denotes that the random variable $s$ has a circularly symmetric complex normal distribution with zero mean and $\sigma^2$ variance.

\section{System model \label{sec:system} }

We consider a wireless distributed  computing system based on a MapReduce framework (cf.~\cite{DGmapreduce:2004, LMYA:17}), where $K$ nodes (servers) first compute Map functions to generate intermediate values for $N$ input files, and then exchange (Shuffle) information over a  wireless interference channel, and finally compute $Q$ outputs (Reduce functions), for some  $K, N, Q \in \mathbb{N}^+$,  with $N\geq K$. The formal model is described as follows.

{\bf{\emph{Map phase}}}: Consider a total of $N$ independent input files $w_1, w_2, \cdots, w_N$. 
Let $\Mc_k \subseteq [1:N]$ denote the indices of the files assigned at Node $k$, $k \in [1: K]$. 
For each file $w_{n},  n \in  \Mc_k$, after the Map function Node~$k$ generates $Q$ intermediate values, i.e., $\{a_{q, n}\}_{q=1}^Q$,  $a_{q,n} \in \mathbb{F}_2^B$, for some $B \in \mathbb{N}^+$.
The  \emph{computation load} of the system is defined as the  total number of map functions computed over $K$ nodes, normalized by the total number of independent files, that is, 
\begin{align}
 r\defeq \frac{\sum_{k=1}^K |\Mc_k|}{N}.  \label{eq:defcomp}
\end{align}

{\bf{\emph{Shuffle phase and the interference channel}}}: In the Shuffle phase,  distributed  nodes exchange the intermediate values over a wireless interference channel, in order to compute Reduce functions.  
Let $\mathcal{W}_k$ denote the indices of Reduce functions computed at Node $k$, $k \in [1: K]$. 
Node $k$ needs the set of intermediate values $\{a_{q, n}:  q\in \Wc_k ,  n \in [1:N] \}$.
Note that after the Map phase, Node $k$ already has
\begin{align}
\Pc_k\defeq \{a_{q, n}: q \in [1:Q],   n\in \Mc_k   \}    \label{eq:sip}
\end{align}
for $k \in [1:K]$.  Therefore, it only requires 
\[ \Gc_k \defeq \{a_{q, n}:  q\in \Wc_k ,  n \in [1:N],  n\notin \Mc_k \}.\]

The communication over this interference channel at time~$t$ is modeled as 
\begin{align}
y_k (t) &= \sum_{i=1}^{K}  h_{k,i}  x_{i}(t) + z_{k} (t), \quad  k \in [1: K], \label{eq:misoy}
\end{align}
where $y_k (t)$ denotes the  received  signal at Node~$k$ at time $t$;  $x_k (t)$ is  the transmitted signal of Node~$k$ at time $t$ subject to a power constraint $\E[| x_k(t)|^2] \leq P$,  and  $z_{k}(t) \sim \mathcal{CN}(0, 1)$ denotes the additive white Gaussian noise (AWGN). $h_{k,i} \in  \mathbb{C}$  denotes the  coefficient of the channel from Transmitter~$i$ to Receiver~$k$, assumed to be fixed and known by all the nodes\footnote{Although we assume that the channel coefficients are fixed, our result also holds for the setting with time varying channel coefficients. An example is provided in Section \ref{sec:varying_channel}. For simplicity of presentation, we will derive our results for fixed channel coefficients.}, for all $ k,i \in [1: K]$.
We assume that all submatrices of the channel matrix consisting of all the channel  coefficients are full rank. 
We also assume that  the absolute value of each channel coefficient is bounded between a finite maximum value and a nonzero minimum value.
We consider the \emph{full-duplex} communication, where each node can receive and transmit signal at the same time.

In this phase, each node first  employs a random Gaussian coding scheme  (cf.~\cite{CT:06}) to encode each of its generated intermediate values $a_{q,n} \in  \mathbb{F}^B_2$ into a \emph{coded packet} $\tilde{\av}_{q,n} \in  \mathbb{C}^{\tau}$, corresponding to $\tau$ channel uses (called a block), for some integer $\tau$ such that $B = \tau \log P + o(\tau \log P)$.
The rate is $B/\tau \approx \log P$  bits/channel use, equivalent to one degree of freedom (DoF). 
The transmission of all the required coded  packets takes place  over a total of $T$  blocks.
In block $\ell$, a subset of the required packets, denoted by $\Dc_{\ell}$, is delivered  to a subset of receivers whose indices are denoted by  $\Rc_{\ell}$,  with each packet intended for one of the receivers, i.e., $|\Dc_{\ell}| = |\Rc_{\ell}|$,  for $\Dc_{\ell} \cap \Dc_{\ell'} = \emptyset$, $\forall \ell, \ell'  \in [1:T], \ell \neq \ell'$.

Specifically, in block~$\ell$ we consider the  \emph{one-shot linear scheme}. The signal transmitted by  Node~$i$, denoted by $\xv_{i}[\ell] \in \mathbb{C}^{\tau}$,  is a linear combination of the coded  packets $\{\tilde{\av}_{q,n}:  \tilde{\av}_{q,n} \in  \Dc_{\ell},   n \in \Mc_i \}$ generated by Node~$i$, that is,
\begin{align}
\xv_i [\ell] &= \sum_{(q, n ): \ \tilde{\av}_{q,n} \in  \Dc_{\ell}, \ n \in \Mc_i }  \beta_{i, q, n}    \tilde{\av}_{q,n},   \label{eq:misoxv}
\end{align}
where $ \beta_{i, q, n} $ is the beamforming coefficient, for $\ell\in [1:T]$ and $i \in [1:K]$.
Then, the received signal of Node~$k$ at block~$\ell$ takes the following form 
\begin{align}
\yv_k [\ell] &= \sum_{i=1}^{K}  h_{k,i} \xv_{i}[\ell]  + \zv_{k} [\ell],    \quad  \ell \in [1: T], \label{eq:misoyv}
\end{align}
where $\zv_{k} [\ell]\in \mathbb{C}^{\tau}$  denotes the noise vector at Receiver~$k$ (Node~$k$) in block $\ell$, for $k \in [1:K]$. 
In terms of decoding,  Node~$k$ utilizes its side information (the generated  coded packets), i.e., 
\[  \tilde{\Pc}_k \defeq \{\tilde{\av}_{q,n}:     a_{q,n} \in  \Pc_k \}\]
 (see \eqref{eq:sip}),  to subtract the interference from  $\yv_k [\ell]$ using a linear function, denoted as,   
\begin{align}
\Lc_{k,\ell} ( \yv_k [\ell],  \tilde{\Pc}_k ).   \label{eq:rxlf}
\end{align}

The communication in block~$\ell$, $\ell \in [1:T]$, is successful if there exist linear operations as in \eqref{eq:misoxv} and \eqref{eq:rxlf} to obtain 
\begin{align}
\Lc_{k,\ell} ( \yv_k [\ell],  \tilde{\Pc}_k ) = \tilde{\av}_{q,n} +  \zv_{k} [\ell]   \label{eq:rxlfp}
\end{align}
for  $\forall k\in \Rc_{\ell}$ and $\tilde{\av}_{q,n} \in \Dc_{\ell} \cap \{ \tilde{\av}_{q',n'}:  a_{q',n'}\in \Gc_k  \}$.
Because the channel in \eqref{eq:rxlfp} is a point to point AWGN channel and its capacity is roughly $\log P$ bits/channel use, $a_{q,n}$ can be decoded  with vanishing error probability as $B$ increases \cite{CT:06}. 
Note that, in our setting we use the random Gaussian coding scheme to encode each of the intermediate values. In terms of decoding, the  maximum likelihood  (ML) decoding can be used. However, the complexity of the  Gaussian coding  and ML decoding is very high. To reduce the complexity, one could use the low-complexity encoding/decoding method, e.g., lattice-based encoding and decoding \cite{EZ:04}.

{\bf{\emph{Reduce phase}}}:  Node $k$ computes the Reduce function $b_q,  q \in \Wc_k,$ as a function of $(a_{q,1}, a_{q,2}, \cdots, a_{q,N} )$. 
In this work we consider a symmetric job assignment, that is, each node has $Q/K$ number of output functions to compute, for $\frac{Q}{K} \in \mathbb {N}$. 
Specifically,
\begin{align}
|\mathcal{W}_1|=|\mathcal{W}_2|=\cdots=|\mathcal{W}_K| = Q/K,   \label{eq:funcsym}
\end{align}
 and $\mathcal{W}_k \cap \mathcal{W}_j= \emptyset$  for any $k, j \in [1:K]$, $k\not =j$.

 We define the \emph{communication load} of this wireless distributed computing  system as
\[L\defeq \frac{T}{NQ} \] which denotes the normalized  communication blocks used in the Shuffle phase.
In our setting,  the computation load  and communication load pair $(r, L)$ is said  to be achievable if there exists a wireless MapReduce scheme consisting of  Map, Shuffle and Reduce phases under the above one-shot linear assumptions, in which all the intermediate values  can be decoded  with vanishing error probability as $B$ increases.
We also define the \emph{computation-communication function} of this wireless distributed computing system, for a given computation load $r$,  as 
\[L^{*} (r) \defeq  \inf \{L: (r, L) \  \text{is feasible} \}. \]

\section{Main results  \label{sec:mainresult}}

This section provides the main results of this work for the wireless distributed computing system  defined in Section~\ref{sec:system}.
The converse and achievability proofs are presented in Sections~\ref{sec:converse} and~\ref{sec:ach}, respectively.

\vspace{5pt}

\begin{theorem}  \label{thm:timeoneshot}
For the wireless distributed computing system defined in Section~\ref{sec:system}, with the assumption of one-shot linear schemes and a sufficiently large $N$,
  the computation-communication function, $L^*(r)$, is characterized as 
\begin{align}
 L^*(r)  =   \frac{ 1- \frac{r}{K} }{\min\{K,2r\} },  \quad \quad  r \in \{1, 2, \cdots, K\}. \label{eq:optimalLr}
 \end{align}
 
\end{theorem}

\vspace{5pt}

Theorem~\ref{thm:timeoneshot} provides a  fundamental tradeoff between the communication load $L$ and the computation load $r$ for the wireless distributed computing system  defined in Section~\ref{sec:system}.
The achievability of Theorem~\ref{thm:timeoneshot} is based on  a one-shot linear scheme that utilizes the methods of zero-forcing and interference cancellation with side information.
The proposed scheme turns out to be optimal for integer $r$. For non-integer $r$, our converse proof shows that $L^*(r) \ge \frac{ 1- \frac{r}{K} }{\min\{K,2r\} }$; our achievability results can be extended using time-sharing such that the line connecting the adjacent integer points $(r, L^*(r))$ and $(r+1, L^*(r+1))$ is achievable, for any $1 \le r \le K-1$, as plotted in Fig. \ref{fig:results}. When $\frac{K}{2} \le r \le K$, the expression in \eqref{eq:optimalLr} is linear in $r$. Therefore, the expression \eqref{eq:optimalLr} gives the optimal computation-communication function for all integer $r$ for $1 \le r \le K$, and all real $r$, for $\frac{K}{2} \le r \le K$.

From the achievability proof in Section \ref{sec:ach}, Theorem~\ref{thm:timeoneshot} holds when $N$ is a multiple of some $N_0$ that depends on $(K,r)$, or when $N$ is sufficiently large for fixed $K,Q,r$.
Note that, in practice, the dataset to be processed is typically big (big data) for the distributed computing systems. The whole dataset can be partitioned into $N$ files and $N$ can be much larger than the number of servers $K$. Moreover,  $ Q$ is often a small multiple of $K$ \cite{DGmapreduce:2004}. We also assume that $r$ is fixed to ensure bounded computation load.

Since the Reduce functions indexed by $\Wc_k$ need $Q N/K$ intermediate values as inputs and $Q\cdot |\Mc_k|/K$ of them have been cached at Node~$k$,  it implies that the total number of  intermediate values required by  Node~$k$ is $\frac {Q }{K}(N-|\Mc_k|)$.
Therefore, the total number of  intermediate values required to be delivered in the Shuffle phase, denoted as $C_{\text{total}}$, can be expressed as  
\begin{align}
C_{\text{total}} &=\sum_{k=1}^{K} \frac {Q }{K}(N-|\Mc_k|)
= {Q N }(1- \frac{r}{K}) . \label{eq:numbers1}
\end{align}

\vspace{4pt}

\begin{remark} [Uncoded TDMA scheme]  \label{rk:tdma}
In the uncoded TDMA scheme, only one node delivers one (uncoded) intermediate value at each transmission block. 
 From \eqref{eq:numbers1}, the communication load $L$ is expressed as 
\begin{align}
L^{\text{Uncoded-TDMA}} (r)  =  1- \frac{r}{K},  \quad \quad r \in \{1,2,  \cdots, K\}.   \label{eq:untdma}
\end{align}
\end{remark}

\vspace{4pt}

\begin{remark}  [Coded TDMA scheme] \label{rk:ctdma}
In the coded TDMA scheme, one node delivers one \emph{coded} intermediate value at each transmission block. From the result in \cite{LMYA:17},  the  communication load $L$ of this coded TDMA scheme is 
\begin{align}
L^{\text{Coded-TDMA}} (r)  = \frac{1}{r} \cdot \bigl(1-\frac{r}{K}\bigr)  ,  \quad  r \in \{1,2,  \cdots, K\}.   \label{eq:codedtdma}
\end{align}
\end{remark}

\vspace{4pt}

\begin{remark}  \label{rk:comp}
The significant improvement of our scheme compared to   uncoded and coded TDMA schemes is depicted in Fig.~\ref{fig:results}.
Note that, the communication load of the proposed one-shot linear scheme is  $(1- \frac{1}{\min\{K, 2r\}}) \times 100 \% $
lower than that of  uncoded TDMA.
Furthermore,  the communication load of the proposed one-shot linear scheme is  $(1- \frac{r}{\min\{K, 2r\}})\times 100 \% $
lower than that of  coded TDMA.
\end{remark}

\section{Examples  \label{sec:example}}

In the introduction, we saw an example of one-shot linear scheme in the Shuffle phase with $K=Q=N=3$ and $r=2$. 
The scheme exploits the side information for interference cancellation.
In this section, we use two examples to illustrate the proposed one-shot linear schemes in the Shuffle phase. In the first example with $r \ge K/2$, the scheme exploits side information cancellation and zero-forcing, while in the second example  with $r < K/2$, the scheme uses side information cancellation and  \emph{partial} zero-forcing.
We introduce important notations including virtual transmitters,  beamforming vectors and channel coefficient vectors for the virtual transmitters. These notations will be used in our converse and achievablility proofs in Sections~\ref{sec:converse} and~\ref{sec:ach}.

\begin{figure*}
\centering
\includegraphics[width=16.5cm]{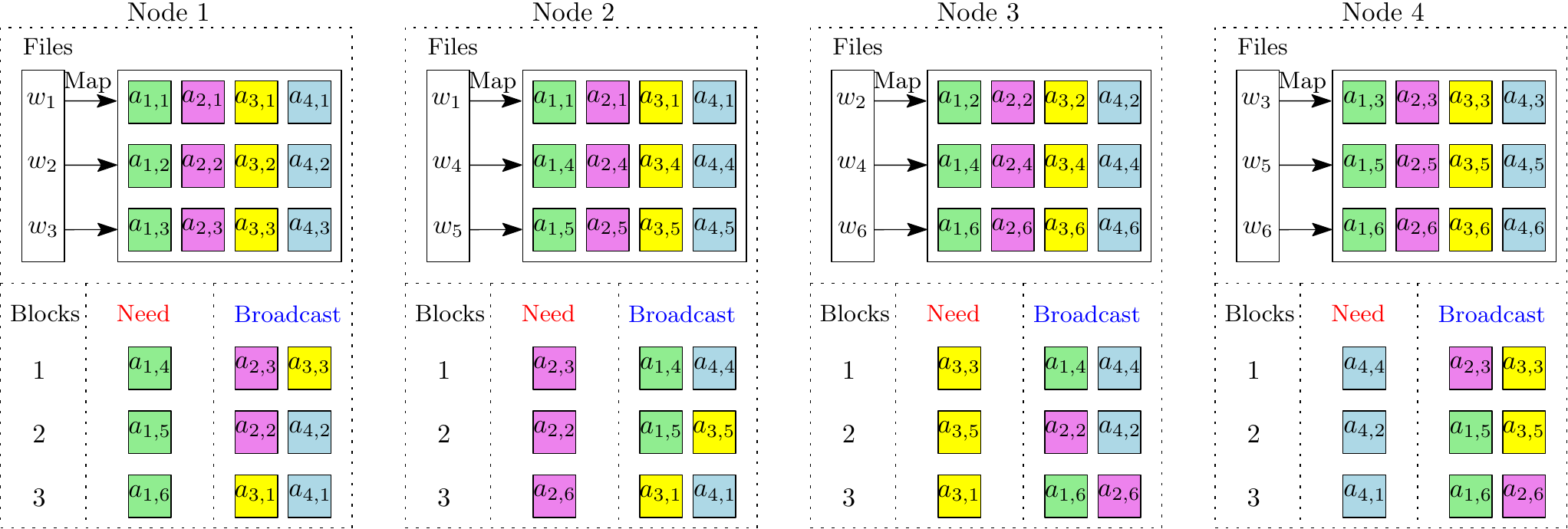}
\caption{An example of wireless distributed computing with $K=Q=4$, $N=6$ and $r=2$.}
\label{fig:KQ4N6}
\end{figure*}

\subsection{The example of $K=Q=4$, $N=6$ and $r=2$ ($r \ge K/2$) }  \label{sec:examplesizf}

Let us consider the case of  $(K=Q=4, N=6, r=2)$. As shown in Fig.~\ref{fig:KQ4N6}, we assign three files  for each node such that $\mathcal{M}_1 = \{1,  2,  3\},  \mathcal{M}_2 = \{1,  4,  5\},   \mathcal{M}_3 = \{2,  4,  6\}$ and $\mathcal{M}_4 = \{3,  5,  6\}$. 
Without loss of generality we consider the case where the $k$-th Reduce function is assigned to Node~$k$, for $k=1,2,3,4$. 

In the Map phase, each node generates a set of intermediate values.
Then, each  intermediate value (e.g., $a_{1,4}$) is mapped into a coded packet (e.g., $\tilde{\av}_{1,4}$).
Let $\mathcal S_n = \{i: n\in \Mc_{i}\}$ represent the indices of all the nodes having file $w_n$, $n\in[1 : N]$. The transmitters indexed by $\Sc_n$ are defined to be a \emph{virtual transmitter} (i.e., virtual Transmitter~$\Sc_n$).
We use 
\begin{align} \label{eq: channel_vector}
{\hv}_{k,\mathcal S_n}\defeq  \big [h_{k, \Sc_n^{1}}, h_{k, \Sc_n^{2}}, \cdots, h_{k, \Sc_n^{|\mathcal S_n|}}\big]^T
\end{align}
 to denote the \emph{channel vector} from virtual Transmitter~$\Sc_n$ to Receiver~$k$, where $\Sc_n^{j}$ denotes the $j$th element of set  $\Sc_n$. 
Let
\begin{align} \label{eq: beamforming_vector}
{\vv}_{{\mathcal S_n},q,n}\defeq \big[ \beta_{\Sc_n^{1}, q, n}, \beta_{\Sc_n^{2}, q, n}, \cdots,\beta_{ \Sc_n^{|\mathcal S_n|}, q, n}\big]^T
\end{align} 
 denote the \emph{beamforming vector} for coded packet $\tilde{\av}_{q,n}$ that is transmitted from virtual Transmitter~$\Sc_n$,  where  $\beta_{\Sc_n^{j}, q, n}$ is the beamforming coefficient of node $\Sc_n^{j}$ for the coded packet $\tilde{\av}_{q,n}$.
For example, for virtual Transmitter $\Sc_n= \{2,3\}$ and Receiver $1$, we have the channel vector ${\hv}^T_{1,\{2,3\}}= \big[h_{1,2}, h_{1,3}\big]$. 
And ${\vv}^T_{{\{2,3\},1,4}} = \big[\beta_{2,1,4}, \beta_{3 ,1,4} \big]$  is the beamforming vector for  the coded packet $\tilde{\av}_{1,4}$. 

In order to compute the first Reduce function, Node~1 needs the intermediate values $(a_{1,1}, a_{1,2}, a_{1,3}, a_{1,4}, a_{1,5}, a_{1,6})$. Since three intermediate values $(a_{1,1}, a_{1,2},a_{1,3})$  are already available at Node~1 after the Map phase,  Node~1 only needs to obtain $(a_{1,4}, a_{1,5}, a_{1,6})$ in the Shuffle phase.  
Similarly,  $(a_{2,2}, a_{2,3}, a_{2,6})$, $(a_{3,1}, a_{3,3}, a_{3,5})$ and  $(a_{4,1}, a_{4,2},a_{4,4})$ need to be delivered to Nodes 2, 3 and 4, respectively (see Fig.~\ref{fig:KQ4N6}).
We will show that in each transmission block, $K=4$ intermediate values are transmitted to $K$ receivers without interference, and three  blocks ($T=3$) are sufficient for delivering all the required intermediate values.

In the first block,  four required intermediate values $a_{1,4}, a_{2,3}, a_{3,3}$ and $a_{4,4}$ are transmitted to Nodes 1,  2, 3 and 4, respectively. Specifically,  the transmitted signals of four nodes  are given as
\begin{align}
\xv_1 [1] &=  \beta_{1, 2, 3} \tilde{\av}_{2,3} + \beta_{1, 3, 3} \tilde{\av}_{3,3},\\
\xv_2 [1] &=  \beta_{2, 1, 4}  \tilde{\av}_{1,4} + \beta_{2, 4, 4}  \tilde{\av}_{4,4},\\
\xv_3 [1] &=  \beta_{3, 1, 4}  \tilde{\av}_{1,4} + \beta_{3, 4, 4}  \tilde{\av}_{4,4},\\
\xv_4 [1] &=  \beta_{4, 2, 3}  \tilde{\av}_{2,3} + \beta_{4, 3, 3} \tilde{\av}_{3,3},
\end{align}
where the beamforming coefficients $\{\beta_{i, q, n} \}$ are  designed  such that   
\begin{align}
\vv_{\{2,3\},4,4} \in   \text{Null}(\hv_{1,\{2,3\}}), &   \  \vv_{\{1,4\},3,3} \in  \text{Null} (\hv_{2,\{1,4\}}),      \label{eq:vec11}  \\
 \vv_{\{1,4\},2,3} \in  \text{Null} (\hv_{3,\{1,4\}}) , & \    \vv_{\{2,3\},1,4} \in  \text{Null} (\hv_{4,\{2,3\}}),     \label{eq:vec22}
 \end{align}
 where $\text{Null} (\ev) $ denotes the null space of the vector $\ev$.

At the receiver side, Node 1 receives the following signal 
\begin{align}
\yv_1 [1] &= \sum_{i=1}^{K}  h_{1,i} \xv_{i}[1]  + \zv_{1} [1] \non \\
&= \underbrace{{\hv}^T_{1,\{2,3\}} \vv_{\{2,3\},1,4} \tilde{\av}_{1,4}}_{\text{desired intermediate value}} + \underbrace{{\hv}^T_{1,\{1,4\}} \vv_{\{1,4\},2,3} \tilde{\av}_{2,3}}_{\text{side information}} +  \non\\
&\quad \underbrace{{\hv}^T_{1,\{1,4\}} \vv_{\{1,4\},3,3} \tilde{\av}_{3,3}}_{\text{side information}} + \underbrace{{\hv}^T_{1,\{2,3\}} \vv_{\{2,3\},4,4} \tilde{\av}_{4,4} }_{\text{interference}}+ \zv_{1} [1]. \non 
\end{align}
In the above expansion of $\yv_1 [1]$, the second and the third  terms can be removed by using side information $\tilde{\av}_{2,3}$ and $\tilde{\av}_{3,3}$ at Node~1,  while  the fourth term can be canceled out due to our design in   \eqref{eq:vec11}. 
In our setting, since we consider the  full rank assumption for the channels, once a beamforming vector is orthogonal to the channel vector associated with the interference, e.g., $\vv_{\{2,3\},1,4} \in  \text{Null} (\hv_{4,\{2,3\}})$, then this beamforming vector is not orthogonal to the channel vector associated with the desired intermediate value, e.g.,  $\vv_{\{2,3\},1,4} \not \in  \text{Null}(\hv_{1,\{2,3\}})$.

Therefore, Node~1 can decode the desired  intermediate value $a_{1,4}$.
Similarly,  Nodes~2, 3 and 4 can decode the desired  $a_{2,3},  a_{3,3}$ and $a_{4,4}$, respectively.

By applying the same methods, in the second block the  desired intermediate values $a_{1,5}, a_{2,2}, a_{3,5}$ and $a_{4,2}$ can be delivered to Nodes~1, 2, 3 and 4, respectively, while in the third  block, the  desired intermediate values $a_{1,6}, a_{2,6}, a_{3,1}$ and $a_{4,1}$  can be delivered to Nodes~1, 2, 3 and 4, respectively. 

Therefore, with the methods of side information  cancellation and zero-forcing, each node can obtain the desired intermediate values after using three blocks ($T=3$) in the Shuffle phase.

\subsection{The example of $K=Q=5$, $N=10$ and $r=2$ ($r < K/2)$}  \label{sec:exampleKQ5N20}

\begin{figure*}
\centering
\includegraphics[width=18cm]{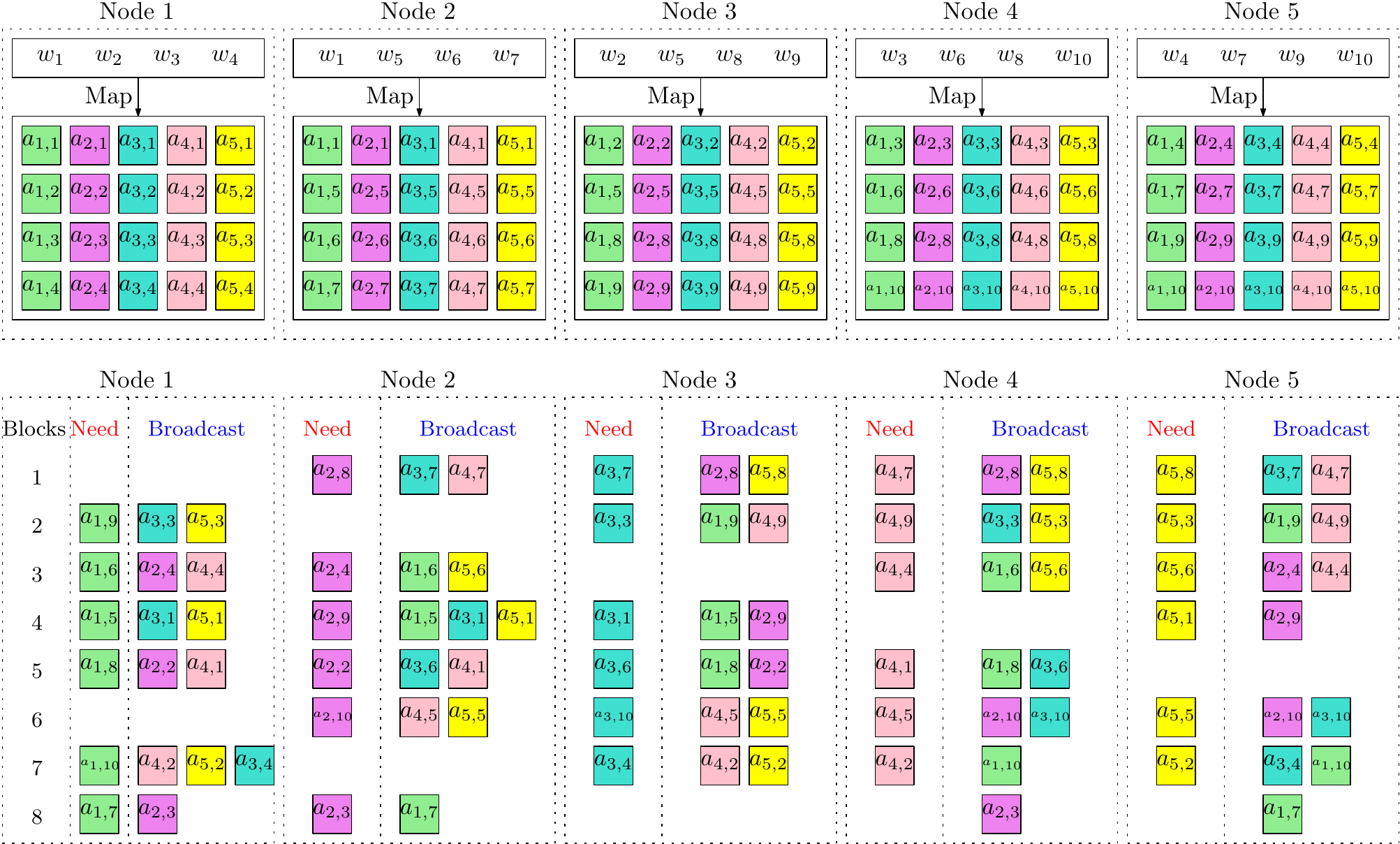}
\caption{An example of wireless distributed computing with $K=Q=5$, $N=10$ and $r=2$.}
\label{fig:KQ5N10}
\end{figure*}

Let us consider the example of  $K=Q=5$, $r=2$ and $N= \binom{K}{r}=10$ (see Fig.~\ref{fig:KQ5N10}). This case is different from the case mentioned in Section~\ref{sec:examplesizf}. In the previous case with  $r \ge K/2$,  $K$ intermediate values are delivered without interference in each transmission block. However, in this case with $r< K/2$, it is impossible to deliver $K$ intermediate values without interference in each transmission block. Instead, $2r$ intermediate values are delivered in each transmission block, by using partial zero-forcing and side information cancellation. 

In this example, given 10 independent files, we assign 4 independent files for each node such that $\mathcal{M}_1=\{1,  2,  3,  4\}$, $\mathcal{M}_2 = \{1,  5,  6,  7\}$, $\mathcal{M}_3=\{ 2,  5,  8,  9\}$, $\mathcal{M}_4=\{ 3,  6,  8,  10\}$, and $\mathcal{M}_5 =\{ 4,  7,  9,  10\}$, as shown in Fig.~\ref{fig:KQ5N10}.  
Again, without loss of generality we consider the case where the $k$-th Reduce function is assigned to Node~$k$, for $k\in [1:K]$. 

After the Map phase, each node generates a set of intermediate values.  In order to complete the computation of each Reduce function, all the nodes need to exchange a subset of  intermediate values in the Shuffle phase. For instance, in order to compute the first Reduce function  at Node 1, the following  intermediate values 
\[(a_{1,5}, a_{1,6},a_{1,7},a_{1,8},a_{1,9},a_{1,10})\] 
need to be delivered to Node 1 in the Shuffle phase. 

We select $2r =4$ nodes to exchange the intermediate values at each transmission block. 
Let us focus on the first block. As shown in Fig.~\ref{fig:KQ5N10}, in this block, we select only  four nodes, i.e., Nodes 2, 3, 4 and 5,  to exchange four intermediate values $(a_{2,8}, a_{3,7}, a_{4,7}, a_{5,8})$. Note that, $a_{2,8}, a_{3,7}, a_{4,7}$ and $a_{5,8}$ are intended for Nodes 2, 3, 4 and 5, respectively.
The beamforming coefficients $\{\beta_{i, q, n}\}$ are designed such that
\begin{align}
\vv_{\{3,4\},5,8} \in   \text{Null}(\hv_{2,\{3,4\}}), &   \  \vv_{\{2,5\},4,7} \in  \text{Null} (\hv_{3,\{2,5\}}),  \label{eq:vec33}  \\
 \vv_{\{2,5\},3,7} \in  \text{Null} (\hv_{4,\{2,5\}}) , & \    \vv_{\{3,4\},2,8} \in  \text{Null} (\hv_{5,\{3,4\}}).  \label{eq:vec44}    
 \end{align}

At the receiver side, Node 2 receives the following signal 
\begin{align}
\yv_2 [1] 
&= \underbrace{{\hv}^T_{2,\{3,4\}} \vv_{\{3,4\},2,8} \tilde{\av}_{2,8}}_{\text{desired intermediate value}} + \underbrace{{\hv}^T_{2,\{2,5\}} \vv_{\{2,5\},3,7} \tilde{\av}_{3,7}}_{\text{side information}} +  \non\\
&\quad \underbrace{{\hv}^T_{2,\{2,5\}} \vv_{\{2,5\},4,7} \tilde{\av}_{4,7}}_{\text{side information}} + \underbrace{{\hv}^T_{2,\{3,4\}} \vv_{\{3,4\},5,8} \tilde{\av}_{5,8} }_{\text{interference}}+ \zv_{2} [1]. \non 
\end{align}
Due to the side information cancellation and zero-forcing, ${a}_{2,8}$ can be decoded at Node~2 without interference. In a similar way,  $a_{3,7}, a_{4,7}$ and $a_{5,8}$ can be decoded at Nodes 3,  4 and 5, respectively.

With the same argument, in each of the other blocks (see Fig.~\ref{fig:KQ5N10}), only four nodes are selected to receive four intermediate values. In this way, at each block all the interference can be either canceled with side information or zero-forced at the selected nodes (\emph{partial} zero-forcing). 
Therefore, all the required intermediate values can be delivered with 8 transmission blocks ($T=8$) in the Shuffle phase. 

Note that it is possible to extend the scheme in this example to accommodate $N=20$ files, and obtain $T=15$ that matches Theorem 1. The details of this extension is shown in Section VI-B.

\subsection{Discussion on time varying channels}
\label{sec:varying_channel}
Note that our achievability and converse  also work  for the setting  with varying channel gains. One simply needs to replace the channel vector and the beamforming vector with the channel matrix and the beamforming matrix, respectively.
 In the following we explain this point by focusing on the  example in Section~\ref{sec:examplesizf}.

For this example with varying channel gains, the received signal of  Node 1 at block~1 takes  the following form 
\begin{align}
\yv_1 [1] &= \sum_{i=1}^{K}  \Hm_{1, i} [1] \xv_{i}[1]  + \zv_{1} [1] \non \\
&= \underbrace{{\Hm}^T_{1,\{2,3\}}  [1] \Vm_{\{2,3\},1,4} \tilde{\av}_{1,4}}_{\text{desired intermediate value}} + \underbrace{{\Hm}^T_{1,\{1,4\}} [1]  \Vm_{\{1,4\},2,3} \tilde{\av}_{2,3}}_{\text{side information}} +  \non\\
&\quad \underbrace{{\Hm}^T_{1,\{1,4\}}[1]  \Vm_{\{1,4\},3,3}  \tilde{\av}_{3,3}}_{\text{side information}} + \underbrace{{\Hm}^T_{1,\{2,3\}}[1]  \Vm_{\{2,3\},4,4}  \tilde{\av}_{4,4} }_{\text{interference}}+ \zv_{1} [1]  \label{eq:matrix}
\end{align}
where 
\begin{align*}
\Hm_{k, i}[\ell] =\left( \begin{array}{cccc}
h_{k, i}^{(1)}[\ell] &  0 & \ldots & 0 \\
0 & h_{k, i}^{(2)}[\ell] & \ldots & 0 \\
\vdots & \vdots &\vdots&\vdots \\
0 & 0 &  \ldots  &h_{k, i}^{(\tau)}[\ell] 
\end{array} \right)
\end{align*}
 and $h_{k, i}^{(n)}[\ell], n\in [1, \tau]$ denotes the channel gain of the $n$-th channel use in block $\ell$, for Transmitter~$i$ and Receiver~$k$.
 In the above expression of $\yv_1 [1]$,  we have the following notations
\begin{align*}
{\Hm}^T_{1,\{2,3\} } [1]= \left( \begin{array}{ccccccc}
h_{1, 2}^{(1)} [1] &h_{1, 3}^{(1)} [1] & 0 & 0 & \ldots & \ldots & 0 \\
0 & 0  & h_{1, 2}^{(2)} [1] &h_{1, 3}^{(2)} [1] &   \ldots & \ldots & 0\\
\vdots & \vdots   & \vdots & \vdots & \vdots & \vdots & \vdots\\
0 & 0  &   \ldots & \ldots & \ldots & h_{1, 2}^{(\tau)} [1] &h_{1, 3}^{(\tau)} [1]
\end{array} \right)
\end{align*}
and 
\begin{align*}
\Vm_{\{2,3\},4,4} = \left( \begin{array}{cccc}
\beta_{2, 4, 4}^{(1)} &0  & \ldots & 0 \\
\beta_{3, 4, 4}^{(1)} & 0  & \ldots & 0\\
0 & \beta_{2, 4, 4}^{(2)} & \ldots & 0\\
0 & \beta_{3, 4, 4}^{(2)} & \ldots & 0\\
\vdots & \vdots  & \vdots & \vdots \\
0 & 0  & \ldots  &\beta_{2, 4, 4}^{(\tau)}\\
0 & 0  & \ldots  &\beta_{3, 4, 4}^{(\tau)}
\end{array} \right)
\end{align*}
where  $\beta_{i, q, n}^{(n) }$ denotes the beamforming coefficient of the $n$-th channel use.
By designing the beamforming coefficients $\{\beta_{i, q, n}^{(n) } \}$ such that  $\Hm_{1,\{2,3\}}^T [1] \Vm_{\{2,3\},4,4}= \mathbf{0} $,
the interference can be removed. 

With this  approach, one can conclude that the proposed general scheme and the converse argument also hold for the setting with time varying channel gains. For simplicity of presentation, we omit the details and just assume fixed channel gains in the remaining sections.

\section{Converse proof for Theorem~\ref{thm:timeoneshot}  \label{sec:converse}}  

In this section we show the converse of Theorem \ref{thm:timeoneshot}.
In fact, we show the following lower bound of the communication load:
\begin{align}
 L^*(r)  =   \frac{ 1- \frac{r}{K} }{\min\{K,2r\} },  \quad \quad  r \in \mathbb{R}, 1 \le r \le K.
\end{align}
We first bound the maximum number of coded packets (of the corresponding intermediate values) that can be transmitted simultaneously in block $\ell$, denoted by $|\Dc_{\ell}|$,  for $\ell \in [1 : T]$. We take a similar approach as in \cite{RLL:12,NMA:17}.
Recall that in block $\ell$ we have coded packets $\Dc_{\ell}$ to be transmitted to the receivers indexed by $\Rc_{\ell}$, with $|\Rc_{\ell}|= |\Dc_{\ell}|$. 
 
In block $\ell$, the transmitted signal from Node~$i$ takes the form as in \eqref{eq:misoxv}. Then, the received signal of Node~$k, k\in \Rc_{\ell}$, takes the following form 
\begin{align}
\yv_k [\ell] &= \sum_{i=1}^{K}  h_{k,i} \xv_{i}[\ell]  + \zv_{k} [\ell]  \non \\ 
&= \sum_{i=1}^{K}  h_{k,i}  \sum_{(q, n ): \ \tilde{\av}_{q,n} \in  \Dc_{\ell}, \ n \in \Mc_i }  \beta_{i, q, n}  \tilde{\av}_{q,n}  + \zv_{k} [\ell]  \non \\ 
&= \sum_{(q, n ): \tilde{\av}_{q,n} \in  \Dc_{\ell}} \sum_{i \in \mathcal S_n}  h_{k,i} \beta_{i, q, n}  \tilde{\av}_{q,n}  + \zv_{k} [\ell]  \non \\
&=  \sum_{(q, n ): \tilde{\av}_{q,n} \in  \Dc_{\ell}} {\hv}^T_{k,\mathcal S_n} \vv_{{\mathcal S_n},q,n} \tilde{\av}_{q,n}  + \zv_{k} [\ell]  \label{eq:eq1}
\end{align}
where the channel vector ${\hv}_{k,\mathcal S_n}$, the beamforming vector ${\vv}_{{\mathcal S_n},q,n}$ are defined in \eqref{eq: channel_vector} and \eqref{eq: beamforming_vector}, respectively.
From (\ref{eq:eq1}), we can conclude that the channel of packet transmission can be transformed into a MISO interference channel.   
The MISO interference channel has $|\Rc_{\ell}|$ single-antenna  receivers and $|\Dc_{\ell}|$ virtual transmitters, where virtual Transmitter~$\Sc_n$  has $|\mathcal S_n|$ antennas, for $n\in [1:N]$. 

In what follows let us first consider the case where each file $w_n$  is stored at $|\mathcal{S}_n|=r$ nodes (symmetric file replications), for  $n = 1,2, \cdots, N$ and integer $r \in \{1,2, \dots, K\}$.  For the other case where different files may be replicated  different times (asymmetric file replications), the proof is provided in Section~\ref{sec:converseasy}.

Let us focus on the transmission of one coded packet $\tilde{\av}_{q,n}$ associated with the intermediate value $a_{q,n}$, for a given pair $(q, n)$. Assume it is transmitted in block $\ell$, and is intended for Receiver $k$, for $\ell \in [1 : T]$ and $k\in [1:K]$. 
Based on a MISO interference channel,  a beamforming vector $\vv_{\Sc_n,q,n}  \in \mathbb{C}^{|\mathcal S_n|}$ is used by  virtual Transmitter~$\Sc_n$ to transmit the corresponding coded packet $\tilde{\av}_{q,n}$.
At the receiver side,  let $\Jc_n = \mathcal{R}_\ell \backslash \{\{k\} \cup \Sc_n\}$ denote the indices of receivers excluding the intended Receiver $k$ and the transmitters indexed by $\Sc_n$, where the packet $\tilde{\av}_{q,n}$ should be zero forced. Then 
\begin{align} \label{eq:size_J}
|\Jc_n | \ge |\mathcal{R}_{\ell}| - |\mathcal{S}_n| -1,
\end{align}
and the inequality  holds with equality when $\mathcal{S}_n$  is a subset of $\Rc_{\ell}$. 
Therefore, for $\Hm \in \mathbb{C}^{|\Jc_n| \times |\mathcal S_n|}$
denoting the channel from virtual Transmitter~$\Sc_n$ to  the receivers indexed by $\mathcal{J}_{n}$,  we should have 
\begin{align}
\Hm \vv_{\Sc_n,q,n} =0      \label{eq:zfcond}
\end{align}
in order to remove  the interference associated with $\tilde{\av}_{q,n}$ at the receivers indexed by $\mathcal{J}_{n}$.  Given that $\Hm$ is full rank and $\vv_{\Sc_n,q,n}$ should be nonzero,  a necessary condition for the existence of the solution to \eqref{eq:zfcond}  becomes 
\begin{align}
|\Jc_n | \leq |\mathcal S_n|-1,
\end{align}
which combined with \eqref{eq:size_J} gives 
\begin{align}
|\Dc_{\ell}| = |\Rc_{\ell}|\leq 2 |\mathcal S_n|=2r.
\end{align}

Furthermore, it is obvious that $|\Dc_{\ell}| \leq K$. Then, we can conclude that, at block $\ell$ the maximum  number of transmitted coded packets satisfies
\begin{align}
|\Dc_{\ell}| \leq \min \{K, 2r\},  \quad \forall \ell \in [1 : T].  \label{eq:maxnum}
\end{align}

Since in one block we can transmit  $|\Dc_{\ell}|$ coded packets, combining (\ref{eq:numbers1}) and (\ref{eq:maxnum}), the number of blocks used to transmit all the intermediate values should be bounded by
\begin{align}
T \geq &\ceil[\bigg]{\frac{C_{\text{total}}}{ |\Dc_{\ell}|}} 
 \ge \ceil[\Bigg]{\frac {NQ (1- \frac{r}{K})}{\min\{K,2r\}}}. \label{eq:symcase2}
\end{align}
Therefore, communication load $L$ should be  bounded by  
\begin{align}
 L = \frac{T}{NQ} \ge \frac{\ceil[\Big]{\frac {NQ (1- \frac{r}{K})}{\min\{K,2r\}}}}{NQ}
\ge \frac{ 1- \frac{r}{K} }{\min\{K,2r\} }.    \label{eq:caser}
\end{align}

\subsection{The case with asymmetric file replications  \label{sec:converseasy} }

Now, let us consider the case  where different files may be replicated different times (asymmetric file replications), given an average computation load $r = \frac{\sum_{k=1}^K |\Mc_k|}{N}$. Note that for this case the value $r$ does not need to be an integer.
Let \[\theta_n  \defeq |\Sc_n|\] denote the  number of times that File $n$ is replicated across the distributed nodes, $n \in [1: N]$. By our definitions of $\theta_n$ and $r$, we have,
\begin{align*}
\frac{\sum_{n=1}^{N} \theta_n}{N} =r.
\end{align*} 
Without loss of generality,  we consider the case with \[\theta_1 \le \theta_2 \le \dots \le \theta_N.\]
Let $C_n $ denote the  total number of  intermediate values generated by File~$n$ and  required to be delivered in the Shuffle phase, $n \in [1: N]$. Then, we have 
\begin{align}
C_n = \frac{(K-\theta_n) Q}{K}. \label{eq:total_i}
\end{align}
This is because, for each node that does not have File~$n$, it needs $Q/K$ intermediate values generated by File~$n$ to complete the computation of its output functions;  and the total number of nodes that do not have File~$n$ is $(K-\theta_n)$. 
It is easy to see that 
\begin{align}
\sum_{n=1}^N  C_n = C_{\text{total}},  \label{eq:sumtotal}
\end{align}
where $C_{\text{total}}$ is defined in \eqref{eq:numbers1}.
Let us use the following notations for the ease of our argument:
\begin{align}
\sigma_n \triangleq & \frac{C_n}{\min\{2\theta_n, K\}} = \frac{(K-\theta_n) }{ \min\{2\theta_n, K\}} \cdot \frac{Q}{K} ,  \label{eq:sigmai}
\end{align}
and 
\begin{align}
\sigma_{\text{sum}} \triangleq & \sum_{n=1}^{N} \sigma_n.    \label{eq:sigmasum}
\end{align}
In the rest of the proof, we show that $\sigma_{\text{sum}}$ is a lower bound on the number of required blocks $T$. Thus the converse of Theorem \ref{thm:timeoneshot} follows from bounding $\sigma_{\text{sum}}$.

In each block~$\ell$, packets corresponding to $ |\mathcal{R}_{\ell}| = |\mathcal{D}_{\ell}|$ intermediate values are transmitted, for $\ell\in [1:T]$.  
Let $r_{\ell,j}$ denote the total number of nodes that generate (after the Map phase) the $j$th intermediate value out of these $|\mathcal{D}_{\ell}|$ intermediate values. It implies that $r_{\ell,j} \in \{ \theta_1, \cdots,  \theta_N\}$, for $j\in [1: |\mathcal{D}_{\ell}|]$.  
For example, in block $\ell$, if we transmit  intermediate values corresponding to Files 1,  1,  2 and  3, then we have
$$(r_{\ell,1} , r_{\ell,2} , r_{\ell,3}, r_{\ell,4} ) = (\theta_1, \theta_1, \theta_2, \theta_3).$$
Without loss of generality let \[r_{\ell,1} \le r_{\ell,2} \le \dots \le r_{\ell ,|\mathcal{D}_{\ell}|}.\]
Let $C_{\ell,n}$ denote the total number of intermediate values generated by File~$n$ and delivered in block~$\ell$. By the definitions of $C_{\ell,n}$ and $C_n$, we have
\begin{align}
\sum_{\ell =1}^{T}  C_{\ell,n} = C_n.   \label{eq:ci998}
\end{align}
Moreover,
\begin{align}
 |\mathcal{D}_\ell| = \sum_{n =1}^{N}  C_{\ell,n} .  \label{eq: totaldl}
\end{align}
Thus
\begin{align}
 \sum_{\ell=1}^{T}  |\mathcal{D}_\ell|   
=& \sum_{\ell=1}^{T}  \sum_{n =1}^{N}  C_{\ell,n}  \label{eq: totalcln}\\
=&   \sum_{n=1}^{N}  C_{n} \label{eq:sumtotal2}\\
=&\sum_{n =1}^{N} \frac{  C_{n}  }{ \min\{ 2\theta_n, K  \}  } \min\{ 2\theta_n, K  \} \non \\
=&\sum_{n =1}^{N}  \sigma_n \min\{ 2\theta_n, K  \} ,\label{eq:sumtotalfinal}
\end{align}
where \eqref{eq: totalcln} is from  \eqref{eq: totaldl}; \eqref{eq:sumtotal2} is from \eqref{eq:ci998}; \eqref{eq:sumtotalfinal} is from \eqref{eq:sigmai}.

Normalizing $\sum_{\ell=1}^{T}  |\mathcal{D}_\ell| $ by $\sigma_{\text{sum}}$ (see \eqref{eq:sigmasum}), we then have
\begin{align}
\frac{1}{ \sigma_{\text{sum}}} \sum_{\ell=1}^{T}  |\mathcal{D}_\ell| =& \sum_{n=1}^{N}  \frac{ \sigma_n}{ \sigma_{\text{sum}}}  \min\{ 2\theta_n, K  \} \label{eq:weighted_ave}\\
\le & \frac{1}{N} \sum_{n =1}^{N}    \min\{ 2\theta_n, K  \} \label{eq:weigtht} \\
\le & \min \bigl\{  \frac{1}{N} \sum_{n=1}^N 2\theta_n , K \bigr\} \label{eq:min} \\
= & \min\{  2r , K \} . \label{eq:bound2244} 
\end{align}
Here \eqref{eq:weighted_ave} is the weighted average of the non-decreasing sequence $ \min\{ 2\theta_n, K  \}$, $1 \le n \le N$, with non-increasing weights $ \frac{ \sigma_n}{ \sigma_{\text{sum}}}$, $1 \le n \le N$. But $\frac{1}{N} \sum_{n =1}^{N}    \min\{ 2\theta_n, K  \}$ is the simple average of $ \min\{ 2\theta_n, K  \}$, $1 \le n \le N$, and thus \eqref{eq:weigtht} holds.
In addition, \eqref{eq:min} is due to the property of the minimum function.

Based on \eqref{eq:bound2244}, we have
\begin{align}
 \sigma_{\text{sum}}
 \ge & \frac{ \sum_{\ell=1}^{T}  |\mathcal{D}_\ell| }{ \min\{ 2r, K  \}} \label{eq:bound22551}\\
= & \frac{ \sum_{n =1}^{N}  C_{n} }{ \min\{ 2r, K  \}} \label{eq:bound22552}\\
= & \frac{ C_{\text{total} }}{ \min\{ 2r, K \}}, \label{eq:bound22553}
\end{align}
where  \eqref{eq:bound22551} is from \eqref{eq:bound2244}; \eqref{eq:bound22552} is from \eqref{eq:ci998} and \eqref{eq: totaldl}; \eqref{eq:bound22553} is from \eqref{eq:sumtotal}.

 Furthermore,  by the same argument as  \eqref{eq:maxnum} we get that 
\begin{align}
  |\mathcal{D}_{\ell}| 
\le & \min\{ 2 r_{\ell,1}, K\}  \label{eq:boundl1}
\end{align} 
where $r_{\ell,1}$ is the smallest number in $\{r_{\ell,j } \}_{j=1}^{|\Dc_{\ell}|}$ for block  $\ell$.
On the other hand,
\begin{align}
1 =& \frac{|\mathcal{D}_\ell|}{|\mathcal{D}_\ell|} \nonumber\\
=&\frac{\sum_{n =1}^{N}  C_{\ell,n} }{|\mathcal{D}_\ell|} \label{eq:5947}\\
\ge & \frac{\sum_{n =1}^{N}  C_{\ell,n}}{\min\{ 2r_{\ell,1}, K\}} \label{eq:7733}\\
\ge & \sum_{n :  C_{\ell,n}=0 } \frac{  C_{\ell,n}}{\min\{2\theta_n, K\}} + \sum_{n :  C_{\ell, n} \neq 0 } \frac{  C_{\ell,n}}{\min\{2\theta_n, K\}} \label{eq:1103}\\
= & \sum_{n =1 }^{N} \frac{  C_{\ell,n}}{\min\{2\theta_n, K\}} , \label{eq:8839} 
\end{align}
where \eqref{eq:5947}  is from \eqref{eq: totaldl}; \eqref{eq:7733} results from \eqref{eq:boundl1};
\eqref{eq:1103} is due to the fact that for all $n$ such that   $C_{\ell,n} \neq 0$, we have $\theta_n \in \{r_{\ell,1}, \dots, r_{\ell, |\mathcal{D}_{\ell}|}\}$, and hence $r_{\ell,1} \leq \theta_n$. Thus,

\begin{align}
T = & \sum_{\ell=1}^T 1 \non \\
\ge &\ceil[\Bigg]{ \sum_{\ell=1}^T    \sum_{n =1 }^{N} \frac{  C_{\ell,n}}{\min\{2\theta_n, K\}}} \label{eq:0011}\\
\ge & \ceil[\Bigg]{\sum_{n =1 }^{N} \frac{ C_{n}}{\min\{2\theta_n, K\}}}  \label{eq:6655} \\
=&  \ceil[\big]{\sigma_{\text{sum}}},  \label{eq:2255}
\end{align}
where \eqref{eq:0011} is from \eqref{eq:8839} and the interger property of $T$;  \eqref{eq:6655} is from \eqref{eq:ci998}; $\sigma_{\text{sum}}$ is defined in \eqref{eq:sigmasum}.
Combining \eqref{eq:bound22553} and \eqref{eq:2255},  the total  number of  transmission blocks $T$ can be bounded by
\begin{align}
T \ge  &\ceil[\big]{\sigma_{\text{sum}}} \label{eq:com01} \\ 
 \ge  & \ceil[\bigg]{\frac{ C_{\text{total} }}{ \min\{ 2r, K \}}} \label{eq:com02} \\ 
=& \ceil[\Bigg]{\frac {NQ (1- \frac{r}{K})}{\min\{2r,K\}}},
\end{align}
where \eqref{eq:com01} is from \eqref{eq:2255}; \eqref{eq:com02} is from \eqref{eq:bound22553}; $ C_{\text{total}} $ is defined in \eqref{eq:numbers1}.
Finally,  the communication load $L$ is 
\begin{align}
 L = &\frac{T}{NQ} \non \\  
 \ge &\frac{\ceil[\Big]{\frac {NQ (1- \frac{r}{K})}{\min\{K,2r\}}}}{NQ} \non \\ 
\ge & \frac{ 1- \frac{r}{K} }{\min\{K,2r\} },
\end{align}
which completes the proof.

\section{Achievability proof for Theorem~\ref{thm:timeoneshot}} \label{sec:ach}

\begin{algorithm}
\caption{Achievable MapReduce Scheme}  \label{generalAlg}
\begin{algorithmic}[1]
\Statex \emph{Map Phase:}
\Procedure{File Placement}{}
\State Partition $\widetilde N$ files into ${\widetilde N}/ {K \choose r}$ disjoint groups 
\For{ $i = 1: {\widetilde N}/ {K \choose r}$ }
   \State Place ${K \choose r}$ files indexed by $[(i-1){K \choose r}+1:i{K \choose r}]$ symmetrically across $K$ nodes, with each file placed at $r$ out of the $K$ nodes
\EndFor
\EndProcedure
\Procedure{Map function}{}
\For{ $k = 1:K $ }
\State Node $k$ computes \emph{Map} functions and outputs $a_{q,n}$, $q\in [1:Q]$ and 
$n \in \mathcal{M}_k$
\EndFor
\EndProcedure
\Statex{}

\Statex \emph{Shuffle Phase:}
\Procedure{Shuffle}{}
\For{ $\ell = 1:T $ }
\State Deliver $\min\{2r, K\!\}$ intermediate values in  block~$\ell$
\EndFor
\EndProcedure
\Statex{}

\Statex \emph{Reduce Phase:}
\Procedure{Reduce function}{}
\For{ $k = 1:K $ }
\State Node $k$ computes \emph{Reduce} functions indexed by $\Wc_k$
\EndFor
\EndProcedure
\end{algorithmic}
\end{algorithm}

In this section, we provide the achievability proof for Theorem~\ref{thm:timeoneshot}.  
We present our file placement scheme as well as the one-shot linear transmission scheme.
We consider the case when  the number of files, $N$, is sufficiently large\footnote{Note that our result also holds for the case with finite $N$ as long as $N$ can be expressed as $N=(\alpha+1)N_0$, for some  nonnegative integer $\alpha$, where $N_0$ is defined in \eqref{eq:nodef}. }.
Note that for a sufficiently large number of files $N$, we have  \[ \alpha N_{0} < N \leq (\alpha+1) N_{0}\] for some nonnegative integer $\alpha$, where $N_0$ is defined by
\begin{align} \label{eq:nodef}
N_0=\left\{\begin{array}{ll}
\binom{K}{r}, & \textrm{if $r \ge K/2$},\\
&\\
\binom{K-r-1}{r-1}\binom{K}{r}, & \textrm{if $r < K/2$}.
\end{array} \right.
\end{align}
In our scheme, we add the following number of empty files 
\[\Delta=(\alpha+1) N_{0}-N, \quad   0 \leq \Delta < N_{0},\] 
and then the number of input files becomes 
\begin{align}   \label{eq:Ntilda} 
\widetilde{N}=N+ \Delta = (\alpha+1) N_{0}.
\end{align}
Afterwards, for every ${K \choose r}$ files, we design a \emph{symmetric} file placement  such that each file is   placed at $r$ out of the $K$ nodes (see Fig.~\ref{fig:KQ4N6} for example). Then, the same placement  can be copied $\widetilde{N}/\binom{K}{r}$ times to complete the placement of $\widetilde N$ input files. 
Since  communication is not needed  when $r \ge K$, we will just focus on the cases when \[r<K.\]

\begin{algorithm}
\caption{Shuffle Phase}  
\label{codingAlg}
\begin{algorithmic}[1]
\vspace{3pt}
\Statex \emph{Shuffle Phase:}
\Procedure{Shuffle}{}

\Procedure{Encoding}{} 
\vspace{3pt}
\State 1. Choose intermediate values: 
\If {$r \ge K/2$}
	\For{block index $\ell  = 1:T$}
		\State	For every $k \in [1:K]$, choose one undelivered $a_{q,n}$ from $\mathcal{G}_k$ as in \eqref{eq:required_by_k}.
	\EndFor
\Else  \hspace{0.5em}($r < K/2$)
	\State Initialize block index $\ell=1$
	\For{every $\Rc \subseteq [1:K]$}
		\For{$copy=1:(\alpha+1)\frac{Q}{K}$}	
			\For{$i = 1: \binom{2r-1}{r}$}
				\State Choose one undelivered $a_{q,n}$ from $A_{k, \Sc_{k,i}}$ defined in \eqref{eq:intermediate_sets} and \eqref{eq:ordered_subsets}, 
				\State  for every $k \in \Rc$.
				\State Increase block index $\ell = \ell +1$.
			\EndFor
		\EndFor
	\EndFor
\EndIf
\vspace{5pt}

\State 2. Gaussian coding:  $a_{q,n} \in  \mathbb{F}^B_2$  $\to$ $\tilde{\av}_{q,n} \in  \mathbb{C}^{\tau}$, where $B = \tau \log P + o(\tau \log P)$, $\forall q, n$.  
\vspace{5pt}

\State 3. Choose beamforming coefficients $\beta_{i, q, n},  \forall q, n, i \in \Sc_n$, to satisfy zero-forcing in \eqref{eq:zfcond}.
\vspace{5pt}

\State  4. Node $i$ sends signal: $\xv_i [\ell] = \sum\limits_{(q, n ): \ \tilde{\av}_{q,n} \in  \Dc_{\ell}, \ n \in \Mc_i }  \beta_{i, q, n}   \tilde{\av}_{q,n}$,  $i \in [1:K], \ell \in [1:T]$.
\EndProcedure
\vspace{5pt}

\Procedure{Decoding}{}
\vspace{5pt}

\State  1. Node $k$ receives signal: $\yv_k [\ell] = \sum_{i=1}^{K}  h_{k,i} \xv_{i}[\ell]  + \zv_{k} [\ell]$, $k \in [1:K], \ell \in [1:T].$
\vspace{5pt}

\State 2. Substract the interference from $\yv_k [\ell]$ by using a linear function, $\Lc_{k,\ell} ( \yv_k [\ell],  \tilde{\Pc}_k )$, where
\State $\tilde{\Pc}_k \defeq \{\tilde{\av}_{q,n}:  a_{q,n} \in  \Pc_k \}$ (side information at Node $k$), $k \in [1:K], \ell \in [1:T]$.
\vspace{5pt}

\State  3. Decode $\tilde{\av}_{q,n}$ as $\Lc_{k,\ell} ( \yv_k [\ell],  \tilde{\Pc}_k ) = \tilde{\av}_{q,n} +  \zv_{k} [\ell]$, $\forall q,n$.
\vspace{5pt}

\State 4. Decoding:  $\tilde{\av}_{q,n} \in  \mathbb{C}^{\tau}$ $\to$  $a_{q,n} \in  \mathbb{F}^B_2$, $\forall q,n$.
\EndProcedure
\EndProcedure
\end{algorithmic}
\end{algorithm}

Similar to \eqref{eq:numbers1}, the total number of intermediate values to be transmitted is  
\begin{align}
\widetilde{N} Q \bigl( 1-\frac{ r}{K} \bigr).   \label{eq:each_receive00}
\end{align}
We describe below the intuition of designing an optimal achievable transmission scheme.
Let us focus on the transmission of one intermediate value $a_{q,n}$, for a given pair $(q, n)$. Assume it is transmitted in block $\ell$, and is intended for Receiver $k$ for $\ell \in[1:T]$ and $k\in [1:K]$. 
Recall that $\Sc_n$ denotes  the indices of  $r$ nodes having the intermediate value $a_{q,n}$. This set of transmitters is viewed as a virtual transmitter. Recall that $\mathcal{R}_\ell$ denotes the indices of receivers in block $\ell$.  $\Jc_n = \mathcal{R}_\ell \backslash \{\{k\} \cup \Sc_n\}$ denotes the indices of receivers where the packet $\tilde{\av}_{q,n}$ is zero forced.
 Thus $|\Jc_n| \le |[1:K] \backslash\{ \{k\} \cup \Sc_n \}| = K-r-1$. 
From the analysis in the converse proof in Section \ref{sec:converse}, the number of receivers without interference from $a_{q,n}$, excluding the intended Receiver $k$, is:  
\begin{align}
&\text{[side information cancellation:] } |\Sc_n \cap \mathcal{R}_\ell | \le |\Sc_n| = r, \label{eq:sideinfoed}\\
&\text{[zero-forcing:] } |\Jc_n| \le \min\{ r-1, K-r-1\}, \label{eq:zeroforced}
\end{align}
and the total number of receivers in a block (i.e., $|\mathcal{R}_\ell | $, $\ell \in[1:T]$) is upper bounded by $1 + |\Sc_n \cap \mathcal{R}_\ell | +  |\Jc_n| \le \min \{2r, K\}$.

We will show an optimal scheme such that $|\mathcal{R}_\ell | = \min \{2r, K\}$ for all $\ell$.
In particular, we show that there exists an assignment of the intermediate values to the blocks, such that for every $a_{q,n}$,
the transmitters indexed by $\Sc_n$ are a subset of the receivers indexed by $\mathcal{R}_\ell$ (i.e., $\Sc_n \subseteq \mathcal{R}_\ell$) and hence \eqref{eq:sideinfoed} holds with equality. As a result, \eqref{eq:zeroforced} automatically holds with equality since $|\Jc_n| = |\mathcal{R}_\ell | - 1 - |\Sc_n| = \min\{ r-1, K-r-1\}$.

For a sufficiently large number of files $N$, the algorithm of the general achievable scheme is described in Algorithm~\ref{generalAlg}.  The algorithm of  the Shuffle phase is described in Algorithm~\ref{codingAlg}.  In what follows, we describe the scheme in details for different cases of $r<K$.

\subsection{The case of  $ r \ge K/2$} \label{sec:casek2k2}

In this case  we will show that $K = \min\{2r, K\}$ intermediate values can be transmitted in each block.
From \eqref{eq:nodef}, in this case we have the following number of data files \[\widetilde{N}=(\alpha+1) N_{0}=(\alpha+1) \binom{K}{r}.\] 
Recall that after the Map phase,  the following set of intermediate values are cached at Node $k$, $k\in [1:K]$,
\begin{align}
\Pc_k\defeq \{a_{q, n}: q \in [1:Q],   n\in \Mc_k   \}   ,
\end{align}
with $|\Pc_k|= Q \cdot |\Mc_k|$,
where $|\Mc_k|= \frac{\widetilde{N}r}{K}$ according to our placement.
Furthermore, the following set of intermediate values are required by Node $k$  
\begin{align} \label{eq:required_by_k}
 \Gc_k \defeq \{a_{q, n}:  q\in \Wc_k ,  n \in [1:\widetilde{N}],  n\notin \Mc_k \}, 
\end{align}
with $|\Gc_k|= \frac {Q }{K}(\widetilde{N}-|\Mc_k|)=  \frac{{\widetilde{N}}Q(1-\frac{r}{K})}{K}$.

In our scheme, we design 
\begin{align}
T= \frac{{\widetilde{N}}Q(1-\frac{r}{K})}{K} \label{eq:tfork}
\end{align} 
blocks such that in every block each of the $K$ nodes receives one intermediate value without interference. Specifically, in each block we choose one of the undelivered intermediate values arbitrarily from  $\Gc_k $, for all $k \in [1:K]$. As a result,  in each block,  $K$ intermediate values are selected,  each intended for a different receiver. For each selected intermediate value, it interferes with $K-1$ unintended receivers.  However, we note that for any intermediate value $a_{q,n}$, \eqref{eq:sideinfoed} and \eqref{eq:zeroforced} hold with equality, since $\mathcal{R}_\ell = [1:K]$,  $|\Sc_n \cap \mathcal{R}_\ell |= |\Sc_n| = r$, and $|\Jc_n| = K-r-1 = \min \{r - 1, K-r-1\}$. Thus a total of $K = \min\{2r,K\}$ intermediate values can be transmitted in every block.

In our scheme, one  intermediate value in $\Gc_k$, $\forall k \in [1 : K]$, is  delivered at each block.  It implies that the number of blocks to deliver all the required intermediate values  is 
\begin{align*}
T= |\Gc_1| = \cdots  |\Gc_K| = \frac{{\widetilde{N}}Q(1-\frac{r}{K})}{K}, 
\end{align*} 
which can be rewritten as  
\begin{align}
T &= \frac{{N}Q(1-\frac{r}{K})}{K} +\frac{\Delta Q (1-\frac{r}{K})}{K},  \label{eq:ttotal3}
\end{align}
where $0 \leq \Delta < N_0,N_0={K \choose r}$ (see \eqref{eq:nodef} and \eqref{eq:Ntilda}). The second term on the right hand side of \eqref{eq:ttotal3} can be bounded by
\begin{align}
\frac{\Delta Q (1-\frac{r}{K})}{K} < \frac{N_0 Q (1-\frac{r}{K})}{K} = o(N),  \label{eq:oN}
\end{align}
where $o(N)/N$ vanishes when $N$  grows and $Q, K, r$ are kept fixed. As mentioned, such scaling of $N$ is seen in many big data applications.
Therefore, for a large $N$, the communication load $L$ is \[L  = \frac{T}{NQ}=\frac{1-\frac{r}{K}}{K}.\]

\subsection{The case of  $  r <  K/2 $} \label{sec:achsmallr}

In this case, at each transmission block we choose $2r=\min\{2r,K\}$ nodes out of $K$ nodes as receivers, and a subset of them as transmitters. Next, we show that $2r$ intermediate values can be transmitted for each block without interference. 

From \eqref{eq:nodef} and \eqref{eq:Ntilda}, in this case we have the following number of data files  \[\widetilde{N}=(\alpha+1) \binom{K-r-1}{r-1} \binom{K}{r}.\]
For any $k \in [1:K]$ and $ \mathcal{S} \subseteq [1:K] \backslash \{k\}, |\mathcal{S}| = r$, let us define a set of intermediate values as
\begin{align}\label{eq:intermediate_sets}
A_{k,\mathcal{S}} =  \{ a_{q,n}: & \ q\in \Wc_k,   \  n \in \cap_{j \in \Sc} \Mc_{j} \}.\non
\end{align}
By definition, for each intermediate value in $A_{k,\mathcal{S}}$,  it is required by Node~$k$ for its Reduce functions and it is cached in each of the nodes indexed by $\Sc$.    
Note that  due to the symmetric file placement, for every pair $(k, \mathcal{S})$, the number of intermediate values in $A_{k,\mathcal{S}}$ is
\begin{align}
|A_{k,\mathcal{S}}| = \frac{Q}{K}\frac{\widetilde{N}}{\binom{K}{r}} = (\alpha+1) \frac{Q}{K} \binom{K-r-1}{r-1}. 
\end{align}

\begin{table}
\small
\caption{An example for one copy with 3 blocks, for $r=2$ and $\mathcal{R}=\{1,2,3,4\}$.} \label{tb:onecopy}
\begin{center}
\begin{tabular}{ccccc}
\toprule
Receiver & 1 & 2 & 3 & 4 \\
\midrule
block 1 & $A^1_{1,\{2,3\}}$ &  $A^1_{2,\{1,3\}}$ &  $A^1_{3,\{1,2\}}$ &  $A^1_{4,\{1,2\}}$ \\
block 2 & $A^2_{1,\{2,4\}}$ &  $A^2_{2,\{1,4\}}$ &  $A^2_{3,\{1,4\}}$ &  $A^2_{4,\{1,3\}}$ \\
block 3 & $A^2_{1,\{3,4\}}$ &  $A^2_{2,\{3,4\}}$ &  $A^2_{3,\{2,4\}}$ &  $A^2_{4,\{2,3\}}$ \\
\bottomrule
\end{tabular}
\end{center}
\end{table}

Let $\mathcal{R} \subseteq [1:K]$ be the indices of an arbitrary set of $2r$ receivers, $|\mathcal{R}|=2r$. We next design $(\alpha+1)\frac{Q}{K}\binom{2r-1}{r}$ blocks such that in every block, every node whose index is in $\mathcal{R}$ receives one intermediate value without interference. Such blocks can be viewed as  $(\alpha+1)\frac{Q}{K}$ copies, each copy corresponding to  $\binom{2r-1}{r}$  blocks. We describe the transmission for one copy, and without loss of generality we index the corresponding blocks of that copy by $1,2,\dots,\binom{2r-1}{r}$. The transmissions for the other copies are the same.

For every $k \in \mathcal{R}$, let 
\begin{align} \label{eq:ordered_subsets}
\mathcal{S}_{k,1}, \mathcal{S}_{k,2}, \dots, \mathcal{S}_{k,\binom{2r-1}{r}}
\end{align}
 be the subsets of $\mathcal{R} \backslash \{k\}$ in any given order, each subset with size $r$, i.e., $|\mathcal{S}_{k, i}| =r$ for $i =1,2, \cdots,  \binom{2r-1}{r}$.  
These subsets are used as different virtual transmitters for Receiver~$k$.
In the $i$-th block, $1 \le i \le \binom{2r-1}{r}$,  one intermediate value in $A_{k,\mathcal{S}_{k,i}}$ is transmitted, for all $k \in \mathcal{R}$. 
From \eqref{eq:sideinfoed} and \eqref{eq:zeroforced}, when an intermediate value  in $A_{k,\mathcal{S}_{k,i}}$ is transmitted, it can be canceled using side information at $r$ undesired receivers  indexed by $\mathcal{S}_{k,i}$ (because it is cached in the nodes indexed by $\mathcal{S}_{k,i}$); it can be zero-forced at the remaining $r-1=\min\{r-1,K-r-1\}$ undesired receivers. Hence, in block $i$, each of  $2r$ receivers in $\Rc$ gets a desired intermediate value without interference. In addition, over the   $\binom{2r-1}{r}$  blocks, a total of $2r\cdot \binom{2r-1}{r}$   intermediate values are transmitted, where each of them comes from one (and only one) of the sets  $\{A_{k,\mathcal{S}}:  k \in \mathcal{R}, \mathcal{S} \subseteq \mathcal{R} \backslash \{k\}, |\mathcal{S}|=r \}$.

For example, let $r=2, \mathcal{R}=\{1,2,3,4\}$. One copy of the scheme has $\binom{2r-1}{r}=3$ blocks. Some details of one copy are given in Table~\ref{tb:onecopy}.
In Table~\ref{tb:onecopy}, $A^{j}_{k,\mathcal{S}}$ denotes the $j$-th element of set $A_{k,\mathcal{S}}$ for $j \in [1: |A_{k,\mathcal{S}}|]$. We can arbitrarily choose the superscript $j$ as long as the intermediate value has not been sent. In this example,  every transmitted intermediate value can be decoded at the intended receiver without interference.  Note that $\{ 2,3\}, \{2,4\}$ and $\{3,4\} $ are three subsets of $\mathcal{R}\backslash \{1\} $ and we choose  $\mathcal{S}_{1,1}=   \{ 2,3\}, \mathcal{S}_{1,2}= \{2,4\}$ and $\mathcal{S}_{1,3}= \{3,4\}$, corresponding to the column for Receiver~1. One can also permute these three subsets in any other order and have, e.g.,  $\mathcal{S}_{1,1}=  \{2,4\}, \mathcal{S}_{1,2}=  \{ 2,3\}$ and $\mathcal{S}_{1,3}= \{3,4\}$.

Now for every $\mathcal{R} \subseteq [1:K]$ of size $2r$, we proceed as before and create $(\alpha+1)\frac{Q}{K}\binom{2r-1}{r}$ blocks. In every block, exactly $2r$ intermediate values can be transmitted without interference. 
Moreover, the scheme is symmetric, in the sense that   a total of $(\alpha+1)\frac{Q}{K}\binom{K-r-1}{r-1}=|A_{k,\mathcal{S}}|$ intermediate values in $A_{k,\mathcal{S}}$  are transmitted at the end of the scheme, for any $k \in [1:K], \mathcal{S} \subseteq [1:K] \backslash \{k\}, |\mathcal{S}| = r$. 
This can be seen from the following facts: there are  $\binom{K-r-1}{r-1}$ choices of $\mathcal{R}$ that include $k$ and $\mathcal{S}$; for every such $\mathcal{R}$ we create $(\alpha+1)\frac{Q}{K}$ copies; and for every copy we transmit one intermediate value in $A_{k,\mathcal{S}}$. 

One can see an example with $(\widetilde{N}=20, K=Q=5, r=2)$ in Table~\ref{tb:k5r2}. 
Let $\{A_{k,\mathcal{S}}^1\}_{k, \Sc}$ be the set of intermediate values associated with Files 1 to 10. Then, we can focus on these intermediate values only in Table~\ref{tb:k5r2} and obtain a scheme with $(N=10, T=8)$ that is identical to the example (see Fig.~\ref{fig:KQ5N10}) in Section \ref{sec:exampleKQ5N20}.
For example,  the four intermediate values $(a_{2,8}, a_{3,7}, a_{4,7}, a_{5,8})$  in  block 1 of  Fig.~\ref{fig:KQ5N10} correspond to the four intermediate values $(A^1_{2,\{3,4\}}, A^1_{3,\{2,5\}}, A^1_{4,\{2,5\}}, A^1_{5,\{3,4\}}) $   in  block 1 of Table~\ref{tb:k5r2}. One can easily extract the scheme in  Fig.~\ref{fig:KQ5N10} from the scheme in Table~\ref{tb:k5r2}. Specifically,  the transmissions of blocks 1, 2, $\cdots$, 8 in Fig.~\ref{fig:KQ5N10}  are identical to the transmissions of blocks 1, 4,  7, 10, 13, 2, 5, 8 in Table~\ref{tb:k5r2}, respectively.

\begin{table*}
\small
\begin{center}
\caption{An Example with $(\widetilde{N}=20, K=Q=5, r=2)$. For each receiver set $\Rc$, we design $(\alpha+1)\frac{Q}{K}=1$ copy of $\binom{2r-1}{r}=3$ blocks. We list all the transmitted intermediate values and the corresponding receivers in each block. Intermediate value $A_{k,\Sc}^1$ corresponds to files 1 to 10; intermediate values  $A_{k,\Sc}^2$ corresponds to files 11 to 20.} \label{tb:k5r2}
\begin{tabu}{cccccc}
\toprule
Receiver (Node) & 1 & 2 & 3 & 4 & 5\\
\midrule
File &$w_1\ w_2\ w_3\ w_4$ & $w_1\ w_5\ w_6\ w_7$ & $w_2\ w_5\ w_8\ w_9$&  $w_3\ w_6\ w_8\ w_{10}$ &$w_4\ w_7\ w_9\ w_{10}$\\
placement&$w_{11}\ w_{12}\ w_{13}\ w_{14}$ & $w_{11}\ w_{15}\ w_{16}\ w_{17}$ & $w_{12}\ w_{15}\ w_{18}\ w_{19}$&  $w_{13}\ w_{16}\ w_{18}\ w_{20}$ &$w_{14}\ w_{17}\ w_{19}\ w_{20}$\\
\midrule
\rowfont{\color{blue}} block $1$  & & ${A^1_{2,\{3,4\}}}$ & $A^1_{3,\{2,5\}}$ &  $A^1_{4,\{2,5\}}$ &  $A^1_{5,\{3,4\}}$ \\
\rowfont{\color{blue}} block $2$   & & $A^1_{2,\{4,5\}}$ &   $A^1_{3,\{4,5\}}$ &   $A^1_{4,\{2,3\}}$ &  $A^1_{5,\{2,3\}}$ \\
 block $3$& & $A^2_{2,\{3,5\}}$ &   $A^2_{3,\{2,4\}}$ &  $A^2_{4,\{3,5\}}$ &  $A^2_{5,\{2,4\}}$ \\
\midrule
\rowfont{\color{blue}} block $4$ & ${A^1_{1,\{3,5\}}}$ & & $A^1_{3,\{1,4\}}$ &  $A^1_{4,\{3,5\}}$ &  $A^1_{5,\{1,4\}}$ \\
\rowfont{\color{blue}} block $5$  & $A^1_{1,\{4,5\}}$ &  & $A^1_{3,\{1,5\}}$ &   $A^1_{4,\{1,3\}}$ &   $A^1_{5,\{1,3\}}$ \\
 block $6$& $A^2_{1,\{3,4\}}$ &  & $A^2_{3,\{4,5\}}$ &  $A^2_{4,\{1,5\}}$ &  $A^2_{5,\{3,4\}}$ \\
\midrule
\rowfont{\color{blue}} block $7$ & ${A^1_{1,\{2,4\}}}$ &  $A^1_{2,\{1,5\}}$ & & $A^1_{4,\{1,5\}}$ &  $A^1_{5,\{2,4\}}$ \\
\color{blue} block $8$  &  \color{blue}$A^1_{1,\{2,5\}}$ &  \color{blue}$A^1_{2,\{1,4\}}$ &  &  $A^2_{4,\{1,2\}}$ &  $A^2_{5,\{1,2\}}$ \\ 
block $9$& $A^2_{1,\{4,5\}}$ &  $A^2_{2,\{4,5\}}$ & & $A^2_{4,\{2,5\}}$ &  $A^2_{5,\{1,4\}}$ \\
\midrule
\rowfont{\color{blue}} block $10$ & ${A^1_{1,\{2,3\}}}$ &  $A^1_{2,\{3,5\}}$ &  $A^1_{3,\{1,2\}}$ & & $A^1_{5,\{1,2\}}$ \\
 block $11$  & $A^2_{1,\{2,5\}}$ &  $A^2_{2,\{1,3\}}$ &  $A^2_{3,\{1,5\}}$ &  &$A^2_{5,\{1,3\}}$ \\
   block $12$ & $A^2_{1,\{3,5\}}$ &  $A^2_{2,\{1,5\}}$ &  $A^2_{3,\{2,5\}}$ &  &$A^2_{5,\{2,3\}}$ \\
\midrule
\rowfont{\color{blue}} block $13$ & ${A^1_{1,\{3,4\}}}$ &  $A^1_{2,\{1,3\}}$ &  $A^1_{3,\{2,4\}}$ &  $A^1_{4,\{1,2\}}$& \\
 block $14$ & $A^2_{1,\{2,3\}}$ &  $A^2_{2,\{1,4\}}$ &  $A^2_{3,\{1,2\}}$ &  $A^2_{4,\{1,3\}}$& \\
  block $15$ & $A^2_{1,\{2,4\}}$ &  $A^2_{2,\{3,4\}}$ &  $A^2_{3,\{1,4\}}$ &  $A^2_{4,\{2,3\}}$ &\\
\bottomrule
\end{tabu}
\end{center}
\end{table*}

Based on the above scheme, and similar to \eqref{eq:ttotal3} and \eqref{eq:oN}, the number of transmission blocks $ T$ is
\begin{align} 
T&=\frac{{\widetilde{N}} Q(1-\frac{r}{K})}{2r}  \label{eq: transblk1}\\
&= \frac{NQ(1-\frac{r}{K})}{2r}+ o(N) \label{eq: transblk3}.
\end{align} 
Finally,  for a large $N$, the communication load $L$ is given as 
 \begin{align}
L  =\frac{T}{NQ}= \frac{1-\frac{r}{K}}{2r}.
\end{align}

\begin{remark} 
As a sanity check, the total number of blocks is also equal to
\begin{align}
T = (\alpha+1)\frac{Q}{K}\binom{2r-1}{r} \binom{K}{2r},  \label{eq: transblk4}
\end{align}
where $(\alpha+1)\frac{Q}{K}\binom{2r-1}{r}$ is the number of blocks for each $\mathcal{R}$, and $\binom{K}{2r}$ is the number of choices of receivers $\mathcal{R}$. One can easily verify that \eqref{eq: transblk4} is equal to \eqref{eq: transblk1}.
\end{remark}

\begin{remark}
In the proof,  at least one copy of $\binom{2r-1}{r}$ blocks for receivers $\mathcal{R}$ is needed. However, it may be possible to reduce the number of blocks in a copy, and hence reduce the minimum required $\widetilde{N}$. The smallest $\widetilde{N}$ for given parameters  is an open problem. 
\end{remark}

\section{Conclusion}   \label{sec:conclusion}

 In this work, we studied the MapReduce-based wireless distributed computing framework, where the distributed nodes exchange information over a  wireless interference network.  
We demonstrated an optimal tradeoff between the computation load and communication load, under the assumption of one-shot linear schemes. One possible future direction is to allow \emph{arbitrary} given file placement in the Map phase, with a given average computation load, and find the corresponding optimal achievable scheme.
Moreover, the communication cost is an open problem when channel state information and synchronization are not fully available. Another direction is to characterize the fundamental  tradeoff between the computation load and communication load without the assumption of one-shot linear schemes, where it may be possible apply the interference alignment approach to improve the system performance.



\end{document}